\documentclass[pdflatex,sn-mathphys-num]{sn-jnl}

\usepackage{graphicx}%
\usepackage{multirow}%
\usepackage{amsmath,amssymb,amsfonts}%
\usepackage{amsthm}%
\usepackage{mathrsfs}%
\usepackage[title]{appendix}%
\usepackage{xcolor}%
\usepackage{textcomp}%
\usepackage{manyfoot}%
\usepackage{tikz}
\usepackage{booktabs}%
\usepackage{algorithm}%
\usepackage{algorithmicx}%
\usepackage{algpseudocode}%
\usepackage{listings}%
\usepackage{adjustbox}
\usepackage{array}
\usepackage{multirow}
\usepackage{color}
\usepackage{colortbl}
\usepackage{subfigure}
\usepackage{xcolor} 
\definecolor{lavender}{rgb}{0.9, 0.9, 0.98}
\usepackage{arydshln}  

\theoremstyle{thmstyleone}%
\theoremstyle{thmstyletwo}%

\theoremstyle{thmstylethree}%

\newcommand{\radial}{\textcolor{blue}{Radial}}
\newcommand{\grid}{\textcolor{purple}{Grid}}
\newcommand{\organic}{\textcolor{teal}{Organic}}

\begin{document}

\title[The Path is the Goal]{The Path is the Goal: a Study on the Nature and Effects of Shortest-Path Stability Under Perturbation of Destination}


\author*[1]{\fnm{Giuliano} \sur{Cornacchia}}\email{giuliano.cornacchia@isti.cnr.it}
\equalcont{These authors contributed equally to this work.}

\author[1]{\fnm{Mirco} \sur{Nanni}}\email{mirco.nanni@isti.cnr.it}
\equalcont{These authors contributed equally to this work.}

\affil[1]{\orgdiv{ISTI}, \orgname{CNR}, \orgaddress{\street{Via G. Moruzzi, 1}, \city{Pisa}, \postcode{56124}, \country{Italy}}}


\abstract{This work examines the phenomenon of path variability in urban navigation, where small changes in destination might lead to significantly different suggested routes. Starting from an observation of this variability over the city of Barcelona, we explore whether this is a localized or widespread occurrence and identify factors influencing path variability. We introduce the concept of ``path stability'', a measure of how robust a suggested route is to minor destination adjustments, define a detailed experimentation process and apply it across multiple cities worldwide. Our analysis shows that path stability is shaped by city-specific factors and trip characteristics, also identifying some common patterns. Results reveal significant heterogeneity in path stability across cities, allowing for categorization into ``stable'' and ``unstable'' cities. These findings offer new insights for urban planning and traffic management, highlighting opportunities for optimizing navigation systems to enhance route consistency and urban mobility.}


\maketitle

\section{Introduction}

Vehicular traffic is one of the critical factors affecting the efficiency of cities, the quality of life of their citizens,  and the environmental impact of transportation.
Nowadays, traffic optimization is of particular relevance, especially in the current context where cities continue to expand in population and density, impacting traffic management, pollution, and road safety. Efficient and sustainable road mobility is therefore essential to ensure that cities remain livable, competitive, and able to meet the needs of their residents. 

\begin{figure*}
\centering
\includegraphics[width=0.68\linewidth]{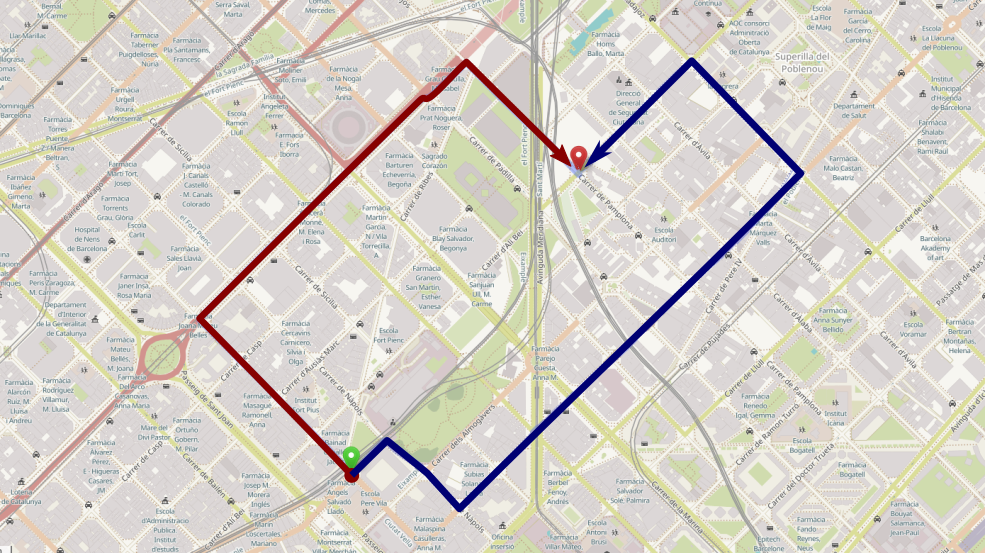}
\caption{Example of two almost identical origin-destination pairs (green and red markers) that lead to different shortest paths (routes generated by OpenStreetMap).\label{fig:motivation}}
\end{figure*}

As urban mobility forms a complex phenomenon, it is extremely important to understand how the travel infrastructures and the city's mobility needs combine, particularly whether the road network provides the correct connections for smooth mass mobility. 
One recent line of research deals with this aspect by assessing the reachability of city destinations in terms of efficient alternative paths available~\cite{washingtonpaper}, which is a measure of road network robustness to high traffic loads and resilience to unexpected events (road closures, accidents, extreme traffic for public events, etc.).
This work focuses on a related topic that stems from an anecdotal observation: in the city of Barcelona, Spain, mobility navigators often provide very different paths to reach two very close destinations, as shown by the example in Figure~\ref{fig:motivation}, where two almost identical destinations are suggested to follow completely different routes. Thus, is this study we try to answer (some of the) several questions that arise: is this variability of paths a peculiarity of Barcelona or a more general phenomenon? What are the features of trips and/or of the city that determine it? Can we recognize patterns in the way this behavior emerges?

To tackle the questions above, we develop an analytical framework implementing the concept of path stability -- namely, a measure of how much small perturbations to the destination of a trip change the best path to reach it -- and then apply it to evaluate empirically the presence and size of the phenomenon in several different cities all around the world, analyzing the results from several perspectives.

Our results show that path stability clearly depends both on city factors, in particular the shape of the road network, and on characteristics of the trips analyzed, such as the distance from the city center. Moreover, there exists a significant variability in the overall path stability across the several cities analyzed, allowing us to talk about \textit{stable} and \textit{unstable} (in terms of path) cities.

This work stems from our previous paper~\cite{cornacchia2024barcelona}, which proposed a first formulation of path instability (here revised) and presented some preliminary results (here greatly extended, including a more robust experimental setting and a much deeper exploration of results).
The source code to reproduce all experiments is available as open source\footnote{\url{https://github.com/GiulianoCornacchia/Shortest-Path-Stability}}.

In the following sections, we review the relevant literature on the topic (Section~\ref{sec:related}), describe our approach (Section~\ref{sec:methods}), present the experiments and results (Sections~\ref{sec:exp_setting} and \ref{sec:results}), and finally provide a discussion and conclusive remarks (Sections~\ref{sec:discussion} and \ref{sec:conclusion}).

\section{Related Work}
\label{sec:related}

Road networks form the backbone of modern transportation systems, serving as the circulatory system that facilitates the movement of people and goods within urban landscapes. Understanding and optimizing traffic flow in these networks is a fundamental challenge in transportation research, with significant implications for efficiency, safety, and environmental sustainability. Despite variations in the geographical features (e.g., seas, rivers, mountains, and coastlines), historical contexts, and socio-economic factors that shape urban areas, road networks exhibit universal characteristics and follow scale-invariant patterns \cite{barthelemy2008modeling}.

\medskip

\textbf{Studying the structure of cities.}
At their core, cities mostly exhibit a monocentric structure, where urban activities are concentrated around a central area, leading to radial expansion. This expansion is often characterized by a density gradient, where population and socio-economic activities diminish with distance from the center \cite{lee2023exploring}. Crucitti et al. \cite{crucitti2006centrality} observed that the road networks of self-organized cities exhibit scale-free properties akin to non-spatial networks, whereas planned cities do not. Additional studies have identified various scaling properties of urban and road networks, such as topological and geographical distances and the distribution of betweenness centrality in urban road networks \cite{lammer2006scaling, chan2011urban, kalapala2006scale, lee2023exploring}.

Urban road networks are almost planar \cite{lammer2006scaling, barthelemy2008modeling} and characterized by a narrow range of node degrees \cite{lammer2006scaling}. Barthelemy et al. \cite{barthelemy2008modeling} identified a power-law relationship between the total network length and the number of nodes. Lee et al. \cite{lee2023exploring} examined the travel-route efficiency of road networks using the detour index (DI) \cite{levinson2009minimum, lee2017morphology}, which measures the ratio between the shortest path distance and the straight-line distance between two locations. Their research revealed universal properties across cities, such as a constant DI regardless of distance from the city center and a negative correlation between DI and road density.

\medskip

\textbf{Routing over the road network.}
The understanding of general road network characteristics has driven extensive research into how to navigate these networks efficiently. While the optimal path (e.g., the shortest or fastest path) represents the most intuitive approach for routing from origin to destination \cite{wu2012shortest, cornacchia2023effects}, relying exclusively on these paths can exacerbate traffic congestion and increase CO2 emissions from a collective standpoint \cite{how_bad_is_selfish_routing, cornacchia2022how, cornacchia2023navigation, cornacchia2024navigation}. A way to overcome this problem and to distribute the vehicles more evenly on the road network is to go beyond the optimal path assignment using Alternative Routing (AR) algorithms \cite{li2022comparing} that provide a plurality of alternative paths between an origin and a destination location in a road network. According to Wang et al \cite{wang2014empirical}, 98\% of roads are underutilized, while 15\% of highway-based routes have faster alternatives, demonstrating the potential for more efficient traffic distribution.

Several methods exist for generating alternative routes. Edge-weight-based approaches compute $k$ alternative paths by iteratively recalculating shortest paths while updating the edge weights after each iteration \cite{li2022comparing, penaltydescription, washingtonpaper}. Plateau-based methods, on the other hand, generate alternatives by identifying common branches between the shortest-path trees of the origin and destination \cite{camvit2005choice}. Other AR algorithms generate $k$ paths that satisfy a dissimilarity constraint and a desired property. For instance, Chondrogiannis et al. \cite{KMO} introduced the $k$-Shortest Paths with Limited Overlap ($k$SPLO), seeking to recommend $k$-alternative paths that are as short as possible and sufficiently dissimilar. Additionally, Chondrogiannis et al. \cite{chondrogiannis2018findind} formalized the $k$-Dissimilar Paths with Minimum Collective Length ($k$DPML) providing $k$ paths sufficiently dissimilar while having the lowest collective path length. Hacker et al. \cite{hacker2021most} proposed the $k$-Most Diverse Near Shortest Paths (KMD), which recommends $k$ near-shortest paths (based on a user-defined cost threshold) with the highest diversity.

AR solutions are also employed in Traffic assignment (TA), where vehicle trips are allocated across a road network to reduce congestion and travel time \cite{campbell1950route}. For example, the METIS algorithm \cite{cornacchia2023oneshot} integrates AR, specifically the $k$-Most Diverse Near Shortest Paths algorithm \cite{hacker2021most}, into a one-shot TA process. This method has shown that generating alternative routes can improve traffic distribution and mitigate negative externalities of congestion \cite{cornacchia2023oneshot}.

\subsubsection*{Position of our work}
Our study focuses on an aspect of mobility over road networks closely related to alternative routing.
In contrast to standard approaches, which aim to induce algorithmically some variability of paths over the road network, we study the intrinsic tendency of a network to spontaneously produce diversified routes. 
Specifically, we introduce the concept of shortest path stability to quantify a road network's propensity to either maintain a consistent path or offer significantly different route options for destinations that are only slightly different.

\section{Methods}
\label{sec:methods}

This section outlines the methodology used to assess the stability of shortest paths within a road network. Path stability refers to the degree to which route alternatives remain similar when the destination location is slightly changed.
Our approach to evaluate path stability involves three key stages: \textit{(i)} generating perturbations to simulate variations in trip destinations, \textit{(ii)} calculating the shortest paths for these perturbed destinations, and \textit{(iii)} measuring the similarity between the original and perturbed paths using a weighted Jaccard index.

\subsection{Path Perturbations}

Given an origin-destination (OD) pair $(o,d)$, namely a trip from a starting location $o$ to an endpoint $d$, we assess its \textit{shortest path stability} $S_\Delta(o,d)$ by creating variations of the destination $d$ applying small perturbations of magnitude $\Delta$ (defined below).
These perturbations simulate slight changes in the destination location, offering insights into how consistent the shortest paths remain.

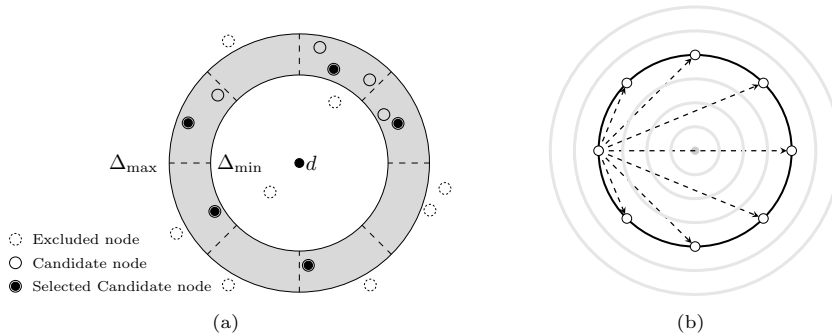
\begin{figure}[htbp]
    \centering
    \subfigure[]{\resizebox{6cm}{!}{
\begin{tikzpicture}
   
        \draw[fill=gray!30, thin] (0,0) circle [radius=2.2cm];
        
        \foreach \angle in {0, 45, 90, 135, 180, 225, 270, 315} {
            \draw[dashed, thin, dash pattern=on 3pt off 3pt] (0,0) -- (\angle:2.2cm);
        }
        
        \draw[fill=white, thin] (0,0) circle [radius=1.5cm]; 

        \filldraw[black] (0,0) circle [radius=0.075cm];
        
        \foreach \angle/\radius in {30/1.65, 80/2, 50/1.85, 140/1.8, 22/1.8, 70/1.7, 160/2, 210/1.65, 275/1.75} {
            \draw[thin] (\angle:\radius cm) circle [radius=0.1cm];
        }
        
        \foreach \angle/\radius in {22/1.8, 70/1.7, 160/2, 210/1.65, 275/1.75} {
            \filldraw[color=black, thin] (\angle:\radius cm) circle [radius=0.07cm];
        }

        \foreach \angle/\radius in {225/.7, 60/1.2, 120/2.4, 350/2.5, 210/2.4, 240/2.4, 300/2.4, 340/2.35} {
            \draw[dashed, thin, dash pattern=on 1pt off 1pt] (\angle:\radius cm) circle [radius=0.1cm];
        }

        \node at (180:1cm) {$\Delta_{\text{min}}$};
        \node at (180:2.8cm) {$\Delta_{\text{max}}$};
        \node at (0:.2cm) {$d$};

        \begin{scope}[shift={(-4.80,-1.3)}]
            \tiny 
            \draw[dashed, thin, dash pattern=on 1pt off 1pt] (0,0) circle [radius=0.1cm];
            \node[right] at (0.2,0) {\footnotesize Excluded node};

            \draw[thin] (0,-0.4) circle [radius=0.1cm];
            \node[right] at (0.2,-0.4) {\footnotesize Candidate node};

            \draw[thin] (0,-0.8) circle [radius=0.1cm];
            \filldraw[color=black, thin] (0,-0.8) circle [radius=0.07cm];
            \node[right] at (0.2,-0.8) {\footnotesize Selected Candidate node};
        \end{scope}
    
    \end{tikzpicture}
    \vspace{.3cm}
}}\hspace{1cm} 
    \subfigure[]{\resizebox{4cm}{!}{
\begin{tikzpicture}
   
        \draw[line width=1.2pt] (0,0) circle [radius=2cm];
        
        \filldraw[gray!40] (0,0) circle [radius=0.075cm];

        \foreach \angleA in {0, 45, 90, 135, 225, 270, 315} {
            \filldraw[color=white, draw=black] (\angleA:2 cm) circle [radius=0.1cm];
        }

        \foreach \angleB in {0, 45, 90, 135, 225, 270, 315} {
            \draw[thin, dashed, ->, >=stealth, line width=0.75pt, shorten >=0.1cm] (180:2cm) -- (\angleB:2cm);
        }

        \filldraw[color=white, draw=black] (180:2 cm) circle [radius=0.1cm];

        \foreach \radiusR in {0.5, 1, 1.5, 2.5, 3}{
         \draw[line width=1.8pt, color=gray!20] (0,0) circle [radius=\radiusR cm];
         }

    \end{tikzpicture}
}}
    \caption{(a) The destination perturbation process for a given destination \(d\). 
     A ring (shaded in gray) of radial boundaries \(\Delta_{\text{min}}\) and \(\Delta_{\text{max}}\), representing the displacement interval, is centered around the destination \(d\). Nodes outside the ring (dashed circles) are excluded. The ring is divided into sectors, each containing candidate nodes (empty circles). For each sector, a representative node (filled circle) is selected based on proximity to the sector's center. (b) The radial sampling approach used for generating origin-destination (OD) pairs, featuring multiple concentric circles centered around the city center. In a detailed view of a single circle (in black), each node is connected to all other nodes along the same circumference, thereby forming several origin-destination pairs. For the sake of readability, here we show sample OD pairs relative to a single origin.}
    \label{fig:diagrams}
\end{figure}

More formally, to perturb the destination location $d$, we identify candidate alternative nodes in the road network within a ring defined by the radial distance interval \(\Delta=[\Delta_{\text{min}}, \Delta_{\text{max}}]\) around the original destination  \(d\) (as illustrated in Figure \ref{fig:diagrams}a). These nodes represent candidate alternative destinations. To ensure even geographic distribution, we divide the ring into $k$ sectors.
For each sector, we select a representative node based on the following criteria: if a sector contains no nodes, no point is selected. If a sector has only one node, that node is selected. In sectors with multiple nodes, we choose the node closest to the center of the sector to maintain spatial balance. This approach results in a set of perturbed destinations $D_{\Delta}(d) = \{d^x\}$, where \(x\) denotes the sector and \(\Delta\) the displacement interval, containing zero to $k$ displaced destinations, depending on node availability.

For an OD pair \((o, d)\), we compute the shortest path \(p(o, d)\) from the origin to the destination. We then calculate the shortest path for each of the perturbed destinations forming the set of perturbed shortest paths \(P_\Delta(o, d) = \{p(o, d^{x}) \mid d^{x} \in D_{\Delta}(d)\}\), where each \(d^{x}\) is a perturbed destination within the displacement interval \(\Delta\).

\subsection{Path Stability}

We evaluate the shortest path stability $S_\Delta(o,d)$ of an OD pair $(o, d)$ by analyzing how much the original shortest path $p(o,d)$ deviates from its perturbed counterparts $P_\Delta(o, d)$ where small changes are introduced to the destination.

We quantify this variation using the Jaccard Index, a widely recognized metric for measuring the similarity between sets. In this context, each route is represented as the set of road segments it traverses.
To account for the relative importance of each road segment, we use a weighted Jaccard Index~\cite{weighted_jaccard_2021}, incorporating road segment lengths as weights. Weighting each segment based on its length ensures that not only the number of shared segments but also their relative significance within the routes is considered.
The weighted Jaccard Index is calculated as follows:

$$
J(A, B) = \frac{\sum_{i \in A \cap B} w_i}{\sum_{i \in A \cup B} w_i}
$$

where $A$ and $B$ are the routes represented as sets of road segments, and $w_i$ is the length of road segment $i$. A value of $J(A, B)=1$ indicates that routes $A$ and $B$ are identical, denoting maximum path stability. Conversely, $J(A, B)=0$ implies that $A$ and $B$ share no common road segments, indicating total path instability. Intermediate values reflect varying degrees of stability, with lower values indicating greater divergence between the routes.

To compute the stability $S_\Delta(o,d)$ for a given OD pair $(o,d)$, we consider the set of perturbed paths $P_\Delta(o,d)$ associated with it. If an OD pair has no perturbed paths, it is excluded from the analysis. When perturbed paths are available, we assess the stability of the shortest path by calculating the Jaccard coefficient between the original shortest path and its perturbed counterparts.
This results in a set of Jaccard values $\left\{ J(p(o, d), \hat{p}) \mid \hat{p} \in P_\Delta(o,d) \right\}$ representing the similarity between the original and perturbed routes. The stability $S_\Delta(o,d)$ of the OD pair $(o,d )$ is then quantified as the average of these Jaccard values:

$$S_\Delta(o,d) = \frac{1}{|P_\Delta(o, d)|} \sum_{\hat{p} \in P_\Delta(o, d)} J(p(o, d), \hat{p})$$

\subsection{City-wide Stability Aggregation}

To assess the overall stability of the shortest paths across an entire city, it is necessary to analyze the shortest path stability of multiple origin-destination pairs within the road network. Given the impracticality of analyzing every possible OD pair, we generate a representative sample of OD pairs that capture the spatial characteristics of the city's road network using a fixed-radius sampling. Finally, we aggregate the path stability values $S_\Delta(o,d)$ for all OD pairs either through global measures or through a box-plot distribution.

\subsubsection{OD Pairs Generation}
\label{sec:pairs_gen}

To systematically sample OD pairs, we employ a fixed-radius sampling method that scans the network radially outward from the city center. This radial sampling approach, commonly used in urban studies to measure spatial metrics or patterns \cite{lee2017morphology, lee2023exploring} avoiding the biases often associated with traditional origin-destination matrices.
As illustrated in Figure \ref{fig:diagrams}b, we drew concentric circles from the city center at $r_{\text{step}}$-km intervals, extending from $r_{\text{min}}$ km to $r_{\text{max}}$ km, to represent increasing distances from the urban core to the periphery. 
For each circle at a given radius, we selected $n$ equally spaced points along the circumference. Each point is matched to the nearest road network node within a given distance threshold. If a point had no nearby match (e.g., if it fell in an inaccessible area such as water bodies, forests, or other unconnected terrain), it is excluded from the sample. The nodes lying on the same circle are paired, generating up to $n\cdot(n-1)$ OD pairs per circle. This systematic approach ensures that the selected OD pairs represent diverse trip lengths and geographic areas, from the urban core to the periphery.

\section{Experimental Settings}
\label{sec:exp_setting}

\begin{figure}
    \centering
    \includegraphics[width=0.95\linewidth]{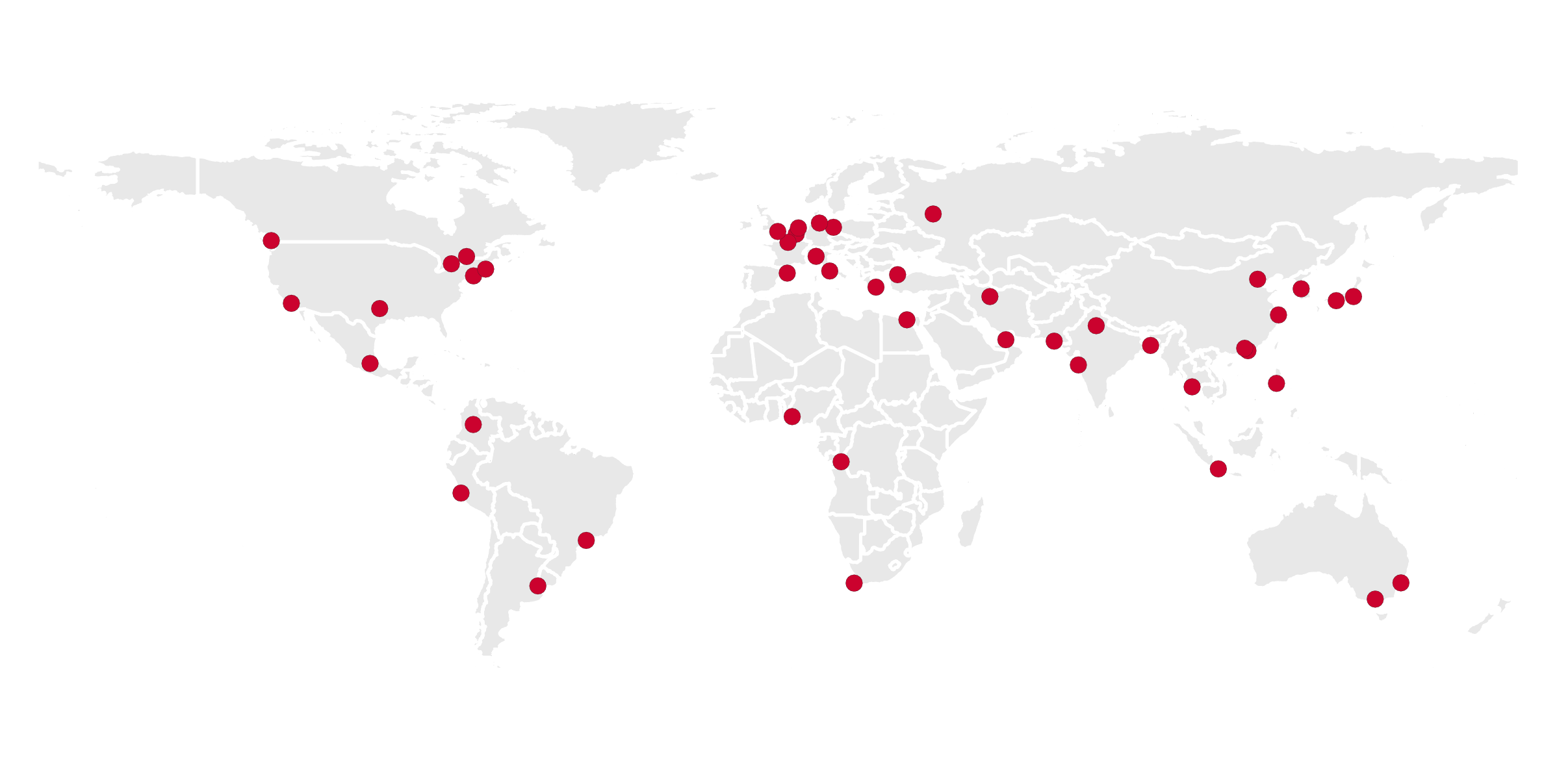}
    \caption{Cities adopted in this work. The selection covers 46 areas from all continents: 16 in Asia, 12 in Europe, 8 in North America and 4 in South America, 4 in Africa and 2 in Oceania.}
    \label{fig:map_cities}
\end{figure}

\textbf{Dataset.}
We compute the path stability for 46 large-scale cities from diverse regions across the globe, shown in Figure~\ref{fig:map_cities}. A complete list is also provided in Table~\ref{tab:divercity_single_cities} on page~\pageref{tab:divercity_single_cities}, together with several indicators that will be described in the next sections. 
The selection includes large cities from all the continents.
The road network data for each geographical area of interest was obtained from OpenStreetMap (OSM\footnote{\url{https://www.openstreetmap.org/}}). \\
\textbf{OD sampling.} 
Following our experimental strategy, for each urban scenario, we generate OD-pairs using the fixed-radius sampling approach. For each city, we identified the city center based on geographic coordinates sourced from \href{latlong.net}{latlong.net}, and from there, we defined concentric circles at radii ranging from $r_{\text{min}}$ = 1 km to $r_{\text{max}}$ = 30 km, with a step size $r_{\text{step}}$ = 1 km. This approach ensures that our OD pairs are distributed across varying distances from the urban core to the periphery, capturing both central and outlying areas. For each concentric circle, we selected 36 evenly spaced points along the circumference at 10° intervals. These points were matched to the nearest road network node within a distance threshold of 500 meters. 
This method generated a maximum of 37,800 potential OD pairs per city, assuming all points were accessible. \\

\textbf{OD perturbation.}
For each OD pair $(o,d)$, we systematically displaced the destination within a radial displacement interval $\Delta = [0, 100$] meters, selecting perturbations in $k$=8 distinct sectors to ensure even geographic distribution around the original destination. 
The choice of the displacement interval was driven by the fact that much larger perturbations will naturally lead to completely different destinations, while our objective is to study the (unexpected) effects of small displacements that do not change the meaning of the trip, mostly keeping at walking distance from the original destination.
In addition, preliminary tests showed that larger displacement intervals yield very similar results, yet with lower stability levels -- as expected.
\\
\textbf{Shortest path computation.}
To retrieve the shortest path between an origin location $o$ and a destination location $d$, we utilize the OSMnx service, which interfaces with OpenStreetMap. Through this service, we obtain a path $p$ representing the shortest route (the one with minimum length) to reach the desired destination $d$, starting from the specified origin location $o$. \\
\textbf{Filtering abnormal perturbed paths.} To ensure realistic perturbed destinations, we exclude any $d^x$ that would require an impractical detour, such as destinations located on the opposite side of or behind the original destination along a highway. Reaching these points would necessitate a U-turn or an exit and re-entry on the highway, making them infeasible for realistic analysis. To identify and exclude these outliers, we calculate the ratio between the length of route $p(d, d^x)$ and the maximum displacement distance ($\Delta_{max}$ = 100 meters). We then remove any $d^x$ where this ratio exceeds the 95th percentile. In our study, this threshold is approximately 30, meaning we exclude any $d^x$ if the length of $p(d, d^x)$ is greater than 3 km.

\section{Results}
\label{sec:results}

In this section, we present the results obtained from the analysis over the 46 large-scale cities described above, based on the computation of shortest path stability for several OD pairs and their comparison against perturbations of the destination locations. 
We try, in particular, to answer a series of basic questions realtive to stability and characteristics of the context: is stability dependent on how close to the city center we are? Is stability variable over different cities? Is there a relation between stability and how much far (and difficult to reach) are perturbed destinations w.r.t. the original ones? Are there features of the road network (e.g. road density or the dominant shape of the network) that impact on stability? Does (in)stability present spatial patterns?

\subsection*{Q1: Is path stability affected by distance from city center?}

Here we study how the perturbation of destinations affects the shortest path across different rings of the radial sampling defined in Section~\ref{sec:methods}.
The results are described in terms of the stability  $S_\Delta(o,d)$ of od-pairs sampled at varying distances from the city center, aggregated across all cities.

\begin{figure}
    \centering
    \includegraphics[width=0.66\textwidth]{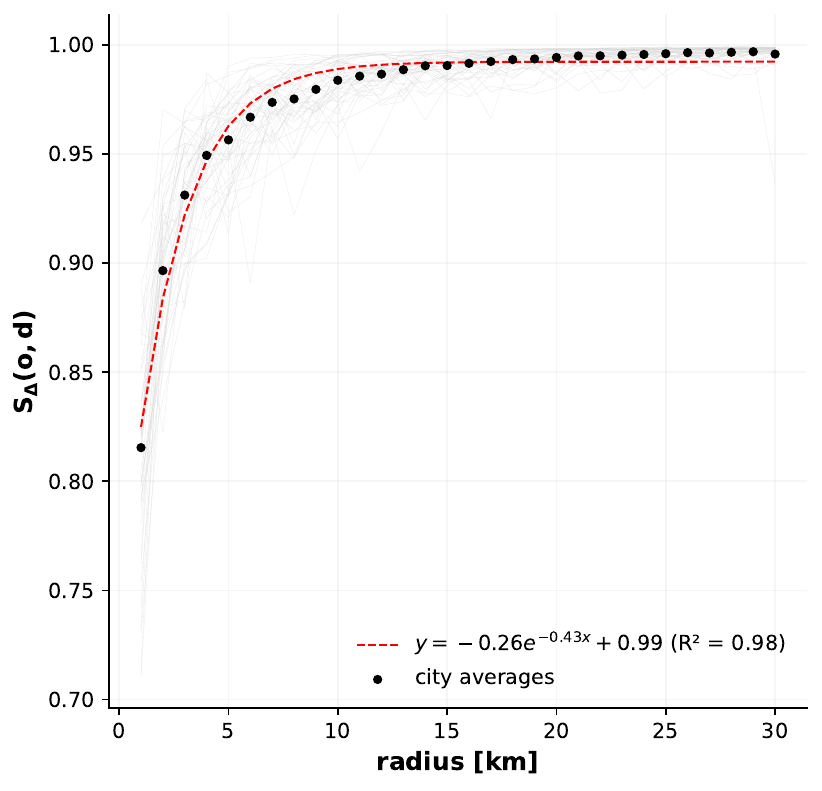}
    \caption{Median path stability $S_\Delta(o,d)$ for OD pairs at varying radial distance $r$ from the city center. Each gray curve represents the stability of a single city, the black points indicate the city averages, and the red line represents exponential fit $\text{$y = a \cdot e^{-bx} + c$}$, where the parameter values are $a=-0.26$, $b=-0.43$, and $c=0.99$, and the $R^2$ value is 0.98.}
    \label{fig:stability_vs_radius}
\end{figure}

Figure \ref{fig:stability_vs_radius} illustrates the relationship between median path stability $S_\Delta(o,d)$ for OD pairs sampled at different distances $r$ from the city center, each gray line representing a different city. The overall trend indicates that path stability increases with distance from the urban core, plateauing at near-complete stability at greater distances.
This trend is well-fitted by the exponential function represented by the equation $\text{$y = a \cdot e^{-bx} + c$}$, where the parameter values are $a=-0.26$, $b=-0.43$, and $c=0.99$, yielding a $R^2$ value of 0.98. The exponential fit highlights that at shorter distances (0-5 km), the median stability values are relatively low, reflecting greater variability in shortest paths when slight perturbations are applied to the destination. As the distance from the city center increases beyond 5 km, the stability values rise sharply. Beyond 10 km, the stability of the paths converges toward 1, meaning that perturbed paths almost entirely overlap with the original shortest paths.
While the general trend is universal across cities, some cities exhibit different levels of stability, as reflected by the slight variations in the individual city curves (see Figure \ref{fig:stability_vs_radius}).

\subsection*{Q2: Which cities are more (un)stable?}

\begin{figure}
    \centering
    \includegraphics[width=1\textwidth]{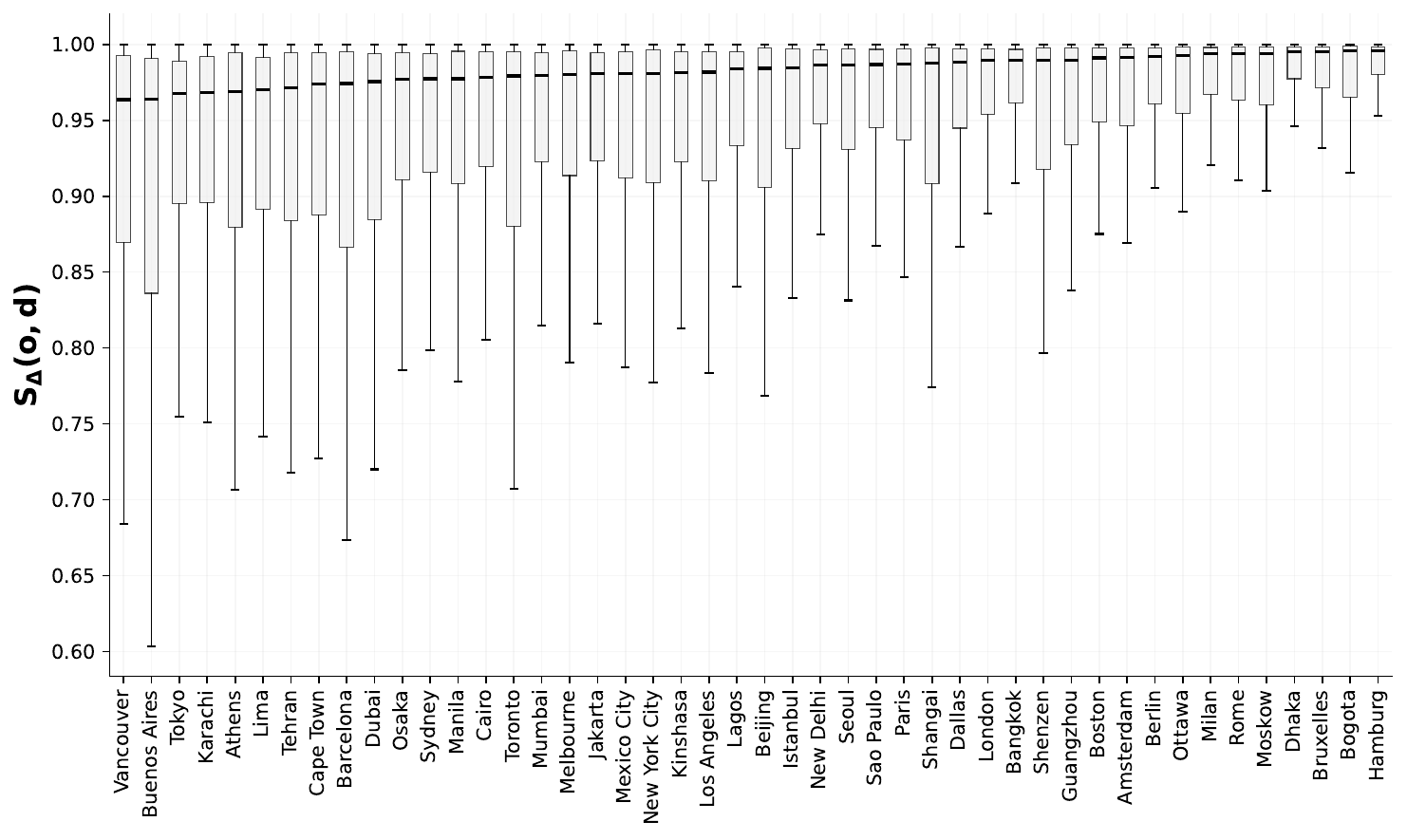}
    \caption{Distribution of path stability $S\Delta(o,d)$ across 46 cities, represented through box-plots sorted in increasing order of median stability.}
    \label{fig:boxplots_s_cities}
\end{figure}

Figure \ref{fig:boxplots_s_cities} presents the distribution of path stability $S_\Delta(o,d)$ among the 46 cities considered in this study through box plots sorted in increasing order of median stability. While the median stability varies only slightly between cities, as expected, the distribution within each city reveals significant heterogeneity. Some cities exhibit a broad range of stability values, while others maintain more consistent stability across their networks.
In general, most cities display high median stability, with values approaching 1, indicating that small perturbations in destination locations typically have minimal impact on the shortest paths. However, variability still exists, as reflected in the spread of the box plots, where the interquartile range highlights the degree of variation within each city. Some cities, such as Hamburg, Dhaka, and Bruxelles, emerge as ``stability islands'', demonstrating consistently high stability for most of the OD pairs and not just on average.
In contrast, cities such as Vancouver, Buenos Aires, and Barcelona—among the least stable—exhibit much wider dispersions and significant variability in path stability across different OD pairs. This suggests that certain areas within these cities experience much higher instability than others. Confirming our intuition, Barcelona ranks among the most unstable cities, placing 8th globally and 2nd in Europe in terms of median path stability and 2nd globally in terms of interquartile range, surpassed only by Buenos Aires.

\subsection*{Q3: Does the magnitude of perturbation affect stability?}

\begin{figure}
    \centering
\includegraphics[width=0.48\textwidth]{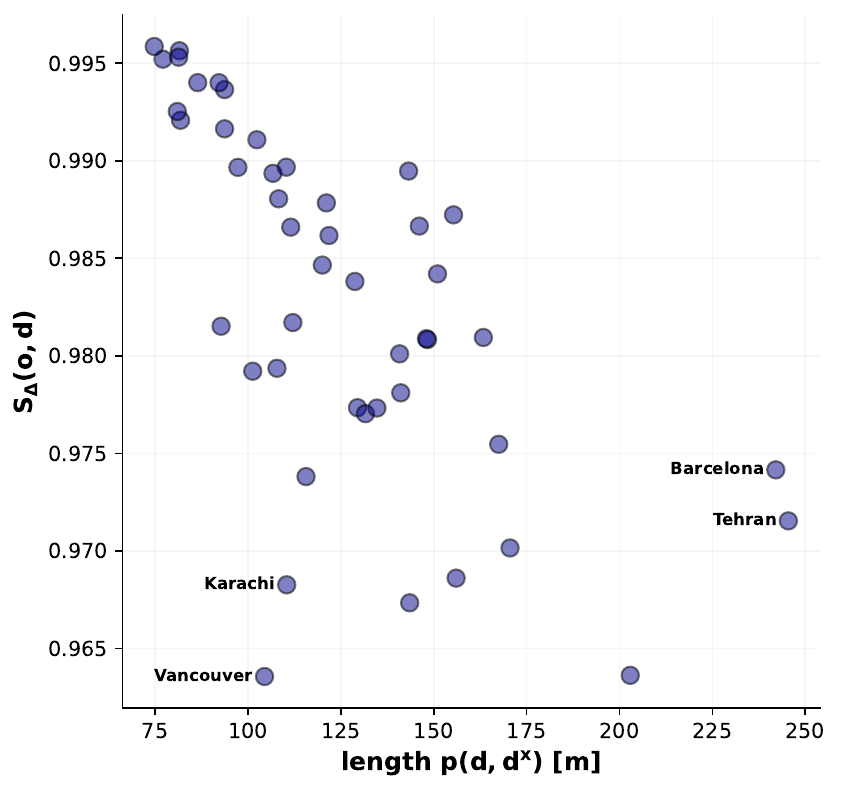}
\includegraphics[width=0.48\textwidth]{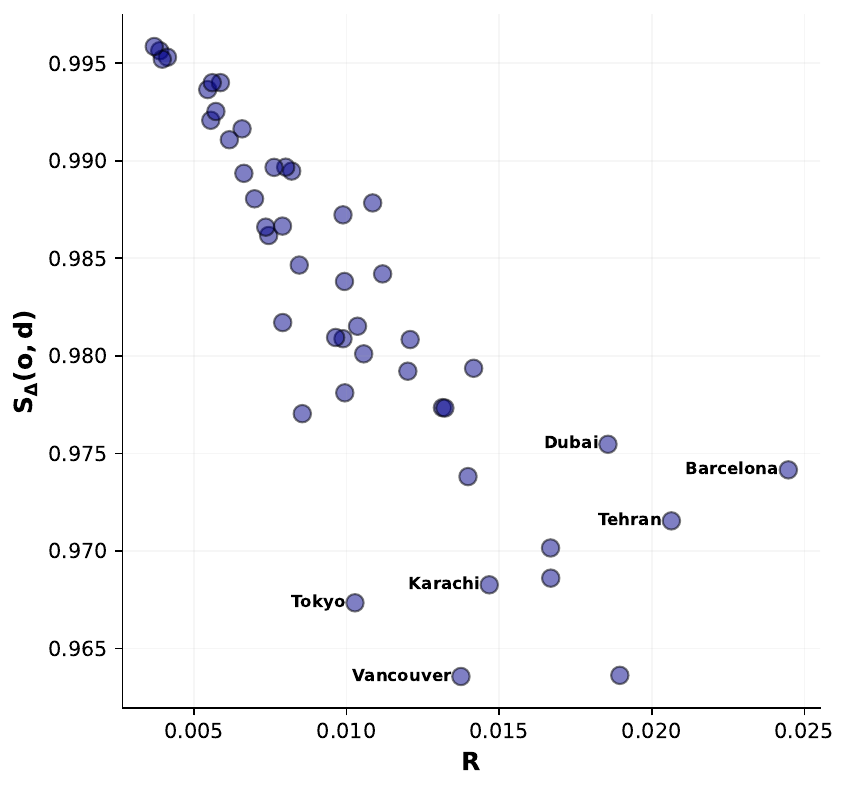}
    \caption{(Left) Shortest path stability $S_\Delta(o,d)$ (median) vs. distance between original destination $d$ and each of its perturbed counterparts $d^x \in D_{\Delta}(d)$ (median). (Right) Same plot, but using a distance $R$ normalized over the original path length.}
    \label{fig:stability_vs_detour}
\end{figure}

To further understand the factors contributing to low path stability, we analyzed the connectivity between each original destination $d$ and each of its perturbed counterparts $d^x \in D_{\Delta}(d)$. 
In particular, we measure the length of the shortest path $p(d,d^x)$ between $d$ and $d^x$ instead of the simple straight line distance, in order to account for the effective travel distance that in some cases can be much larger.
This measure can be interpreted as the effort required to reach the new (perturbed) destination while keeping the path followed to reach the original one. 
Figure \ref{fig:stability_vs_detour}(left) reveals a strong negative correlation (Pearson = -0.634) between the median $p(d,d^x)$ length and path stability, suggesting that cities where perturbed destinations are closely connected (in terms of route length) to the original destination tend to have higher path stability -- or, in other terms, the original shortest path needed only minor corrections to reach the new destination--, while cities with longer $p(d, d^x)$ generally exhibit lower stability.
While this pattern applies to most cities, some outliers deviate from it. For instance, Vancouver and Karachi show relatively low stability (0.964 and 0.968, respectively) despite short $p(d, d^x)$ lengths (105 and 110 meters), indicating notable sensitivity to small destination shifts. Meanwhile, cities like Barcelona and Tehran show both low stability (0.974 and 0.971, respectively) and longer $p(d, d^x)$ paths (242 and 246 meters).

We extend the analysis by normalizing the length of $p(d,d^x)$ w.r.t. the original route length, namely $R = l(p(d,d^x)) / l(p(o,d))$, where $l(x)$ denotes the length of path $x$, in meters.
Figure~\ref{fig:stability_vs_detour}(right) illustrates the relationship between route stability and the ratio $R$.
The negative correlation against stability is confirmed and also significantly strengthened (Pearson = -0.831). 
We can also observe that the normalization moved some outliers of the previous plot close to the general trend (Karachi and Vancouver in the left part, Tehran on the right part), while Barcelona remained far from it, confirming its outstanding behaviour.

\subsection*{Q4: Do features of the road network affect stability?}

The ability of a road network to exhibit stable or unstable alternative paths is inherently shaped by the underlying network frame. To try to unveil the impact of the road network on stability, we investigated the relationship between path stability $S_\Delta(o, d)$ and network-based measures.
We considered: average street length, its standard deviation, road circuity (defined as ratio between road length and the straight line distance of its endpoints), density of intersections and roads, total road length.

\begin{figure}
    \centering
\includegraphics[width=0.48\textwidth]{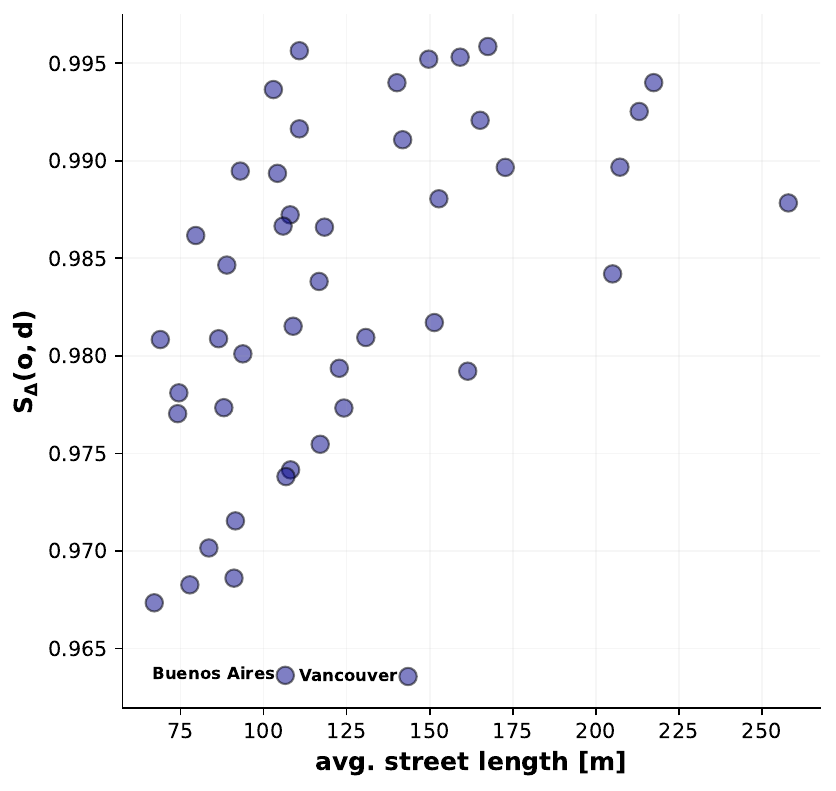}
\includegraphics[width=0.48\textwidth]{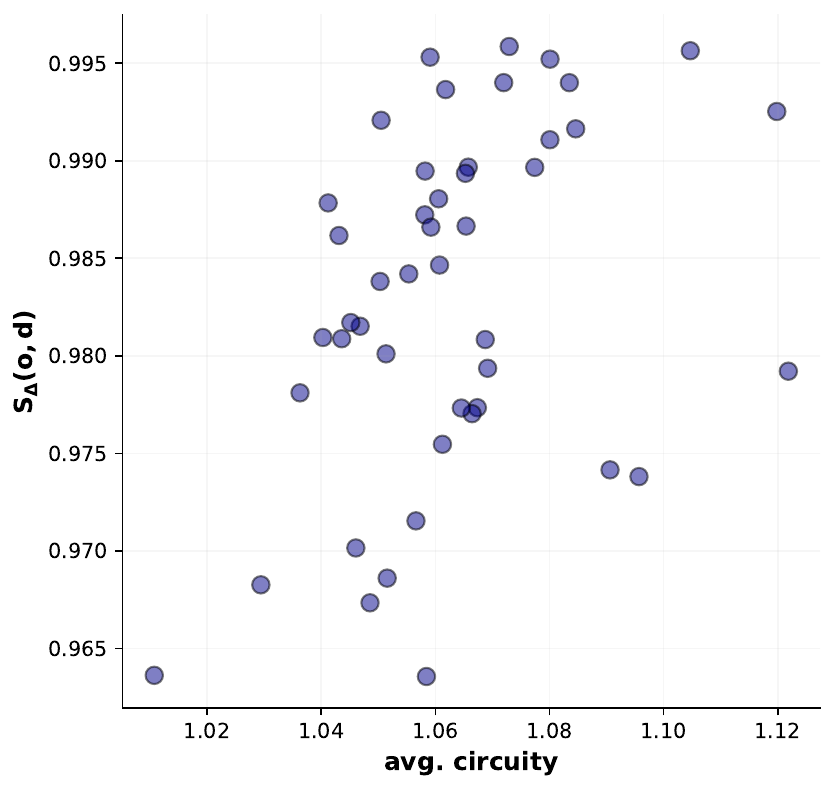}
\caption{Shortest path stability $S_{\Delta}(o,d)$ (median) vs. average street length (Left) and vs. average road circuity (Right). The smaller is the latter, the more straight the roads are.}
\label{fig:stability_vs_elen}
\end{figure}

We found a moderate positive Pearson correlation of 0.476 between path stability and average street length across cities (see Figure~\ref{fig:stability_vs_elen}(left)), indicating that cities with longer average street segments experience higher path stability. Indeed, longer segments reduce intersections, or ``decision points'', minimizing route divergence. In contrast to this general trend, Buenos Aires and Vancouver emerge as outliers, displaying lower-than-expected stability despite longer street segments.
Moreover, we find a moderate correlation between stability and the standard deviation of edge lengths (Pearson = 0.593), suggesting that greater variability in edge length enhances stability. Finally, we found a smaller yet significant correlation with the edge-level circuity (Pearson = 0.401, shown in Figure~\ref{fig:stability_vs_elen}(right)). No meaningful relationships emerged between stability and the other measures.

\subsection*{Q5: Is the shape of the road network relevant?}
Beside statistical properties of the road network, also its geometrical configuration might in principle have a strong impact on how shortest paths change.
Modern, very regular networks typically allow several alternative paths of similar length to reach the same destination, whereas networks in older or geographically constrained cities provide more limited choices.

We characterize such configurations in two ways: 
\begin{enumerate}
    \item \textbf{Entropy of orientation:} we measure how much the road network has a strict grid shape by analyzing the bearing of roads, measuring the entropy of its probability distribution following the methodology outlined in \cite{boeing2019urban}. Grid cities will concentrate bearings along very few directions (for instance four, if all the city grids are aligned), and will thus be identified by lower entropy values, whereas non-grid ones will be more chaotic, and thus yield a higher entropy;
    
    \item \textbf{Urban studies approach:} we refer to the traditional classification adopted in the studies on urban configurations, consisting in three categories: Grid, defined as above; Radial, where roads mainly depart from a center of the city; and Organic, where roads developed around landmarks (hills, rivers, walls, etc.) and/or were added in stratifications without a consistent plan~\cite{2024Designing}. In this case, our cities were associated to the corresponding class through visual inspection of the road network.
\end{enumerate}

\begin{figure}
    \centering
    \includegraphics[width=0.48\linewidth]{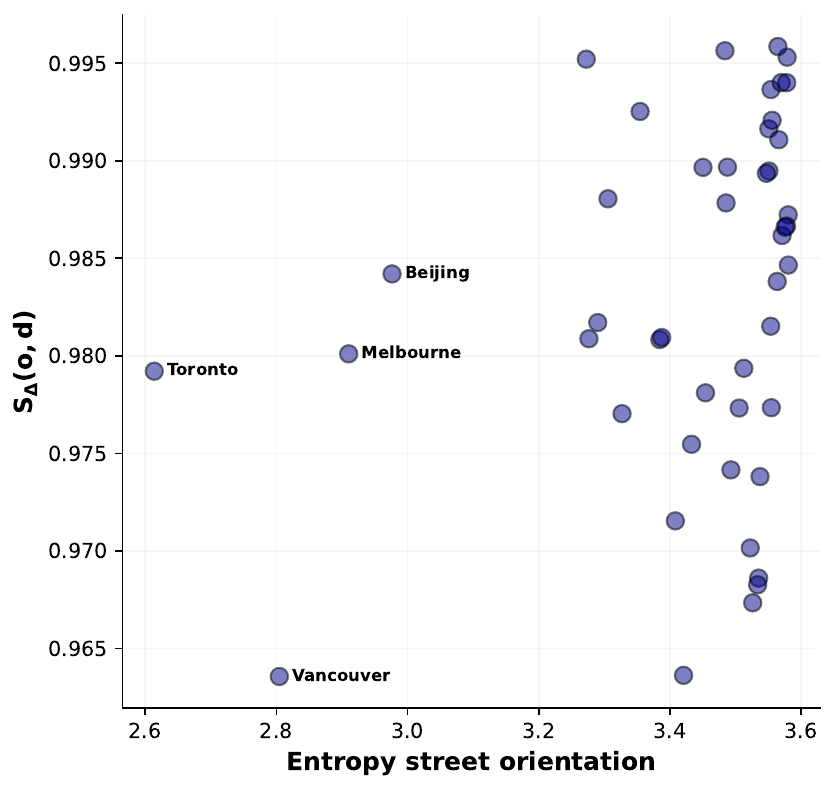}
    \caption{Shortest path stability $S_\Delta(o,d)$ vs. entropy of road bearing (median). Low entropy values represent cities with a grid-like structure.}
    \label{fig:stability_vs_entropy}
\end{figure}

\textbf{Entropy-based categorization.}
Here we study the relation between stability and entropy of road bearing, through a plot similar to the previous ones. 
The results, depicted in Figure~\ref{fig:stability_vs_entropy}, show a weak positive correlation between the two values (Pearson = 0.25). 
In particular, we can identify a large top-right group of cities with high entropy (most likely having an irregular road network) associated to medium-high stability; and a small group of four cities (Toronto, Vancouver, Melbourne and Beijing) showing a very low entropy and medium-low stability. These results provide a first confirmation of the hypothesis that highly regular (i.e. low entropy) road networks in general lead to lower shortest path stability, although results in the cases of medium and high entropy remain a bit unclear.

\textbf{Urban studies classification.} 
In Table~\ref{tab:city_shapes} we associate the road network of the cities under analysis to the three groups mentioned above (Grid, Radial, Organic), also providing, for each city, a thumbnail representation of the road network covering a $6 \times 6$ km area around the center. The most common configuration is the Organic one (27 cities), which is predominant in Europe and Asia, followed by Grid (15 cities), predominant in North America, and only a few Radial ones (4 cities), most of them in Europe.

\begin{table}[h!]
\setlength\tabcolsep{1.5pt}
\centering
\vspace{-0.5cm}
\begin{tabular}{cc m{6em} cc m{6em} cc m{3.5em}}
\textbf{City} & \textbf{Shape} & \textbf{} & \textbf{City} & \textbf{Shape} & \textbf{} & \textbf{City} & \textbf{Shape} & \textbf{} \\
\hline


%
Amsterdam & \radial{} & \adjustbox{valign=m}{\includegraphics[width=0.083\textwidth]{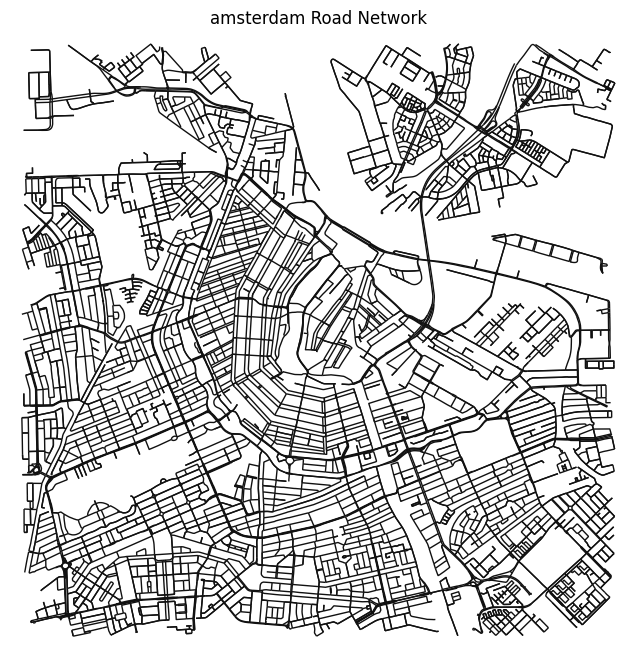}} & 
Athens & \organic{} & \adjustbox{valign=m}{\includegraphics[width=0.083\textwidth]{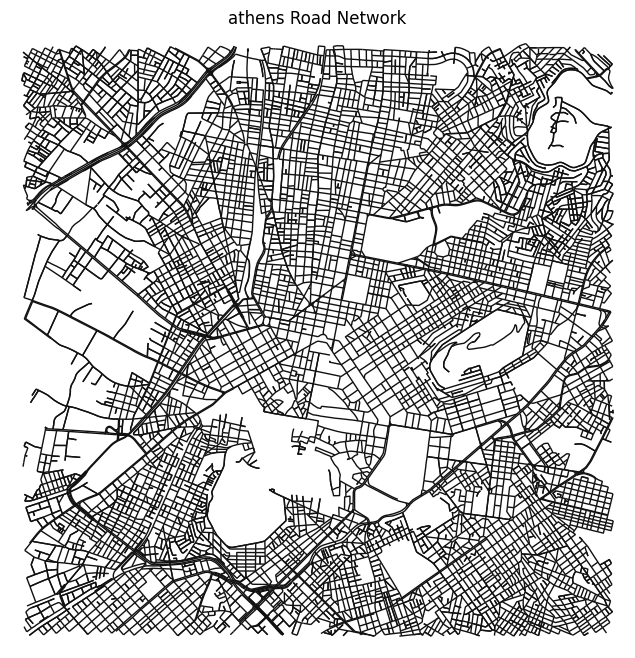}} & 
Bangkok & \organic{} & \adjustbox{valign=m}{\includegraphics[width=0.083\textwidth]{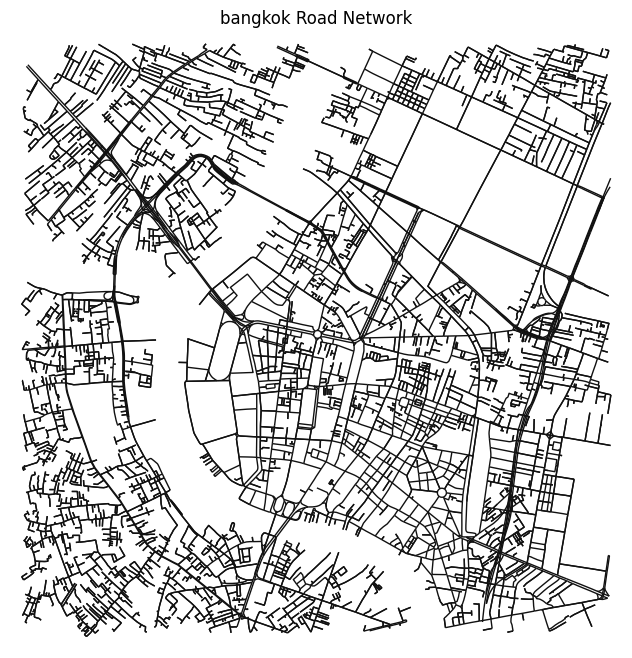}} \\
\hline
Barcelona & \grid{} & \adjustbox{valign=m}{\includegraphics[width=0.083\textwidth]{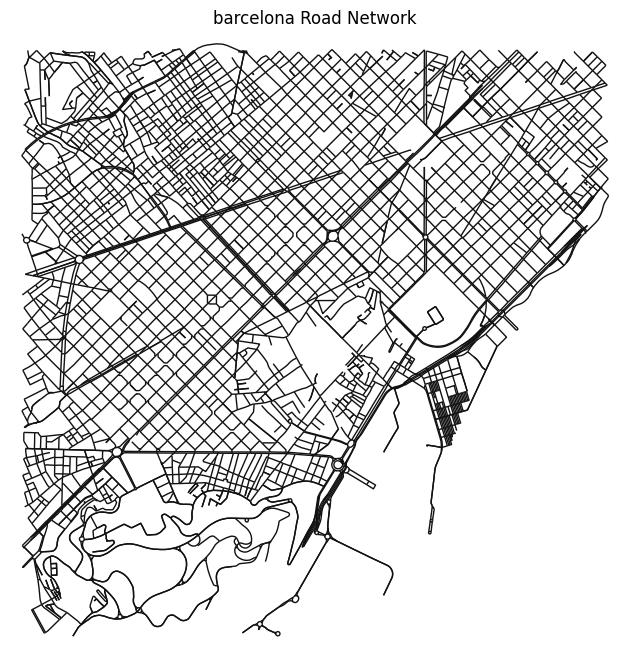}} & 
Beijing & \grid{} & \adjustbox{valign=m}{\includegraphics[width=0.083\textwidth]{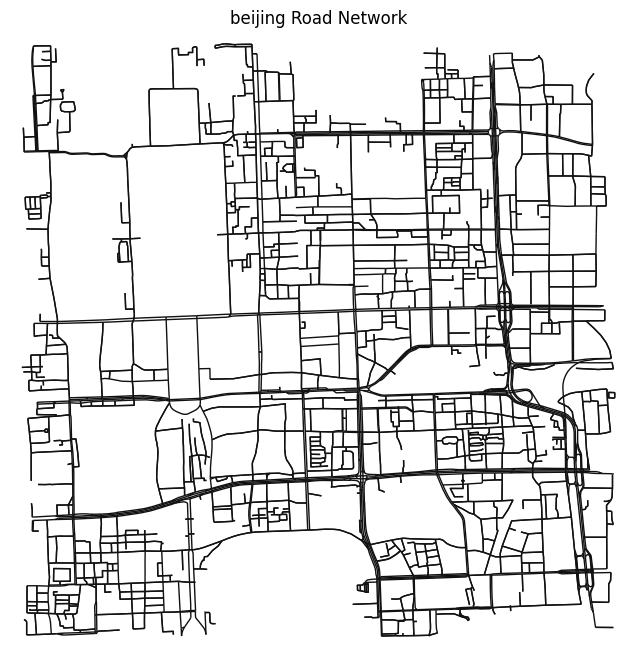}} & 
Berlin & \radial{} & \adjustbox{valign=m}{\includegraphics[width=0.083\textwidth]{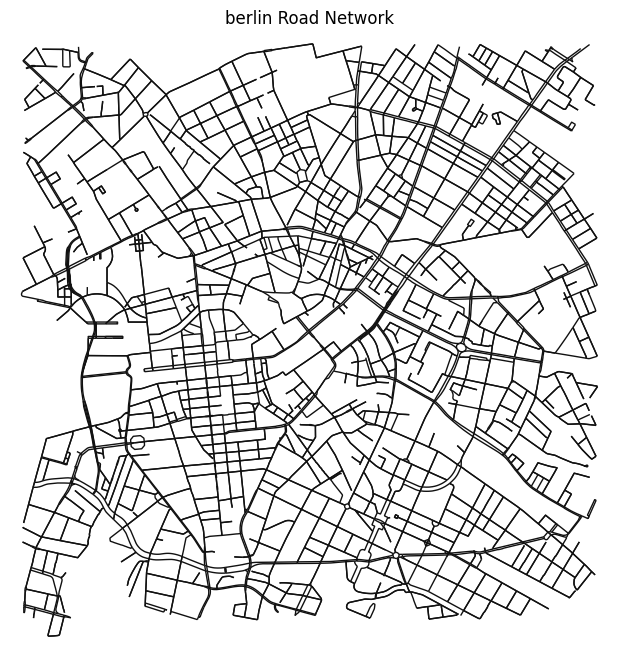}} \\
\hline
Bogota & \grid{} & \adjustbox{valign=m}{\includegraphics[width=0.083\textwidth]{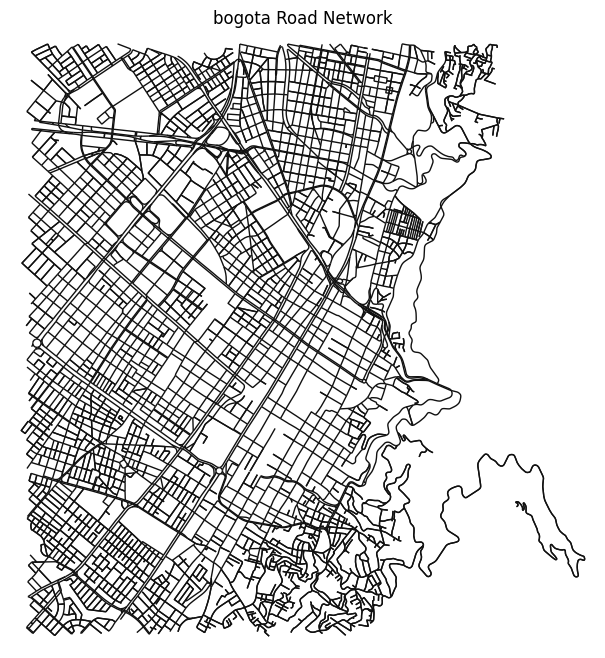}} & 
Boston & \organic{} & \adjustbox{valign=m}{\includegraphics[width=0.083\textwidth]{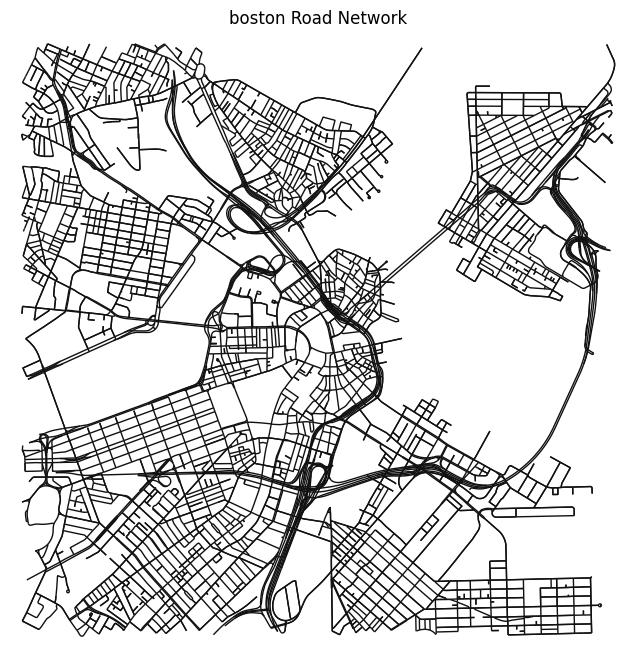}} & 
Bruxelles & \organic{} & \adjustbox{valign=m}{\includegraphics[width=0.083\textwidth]{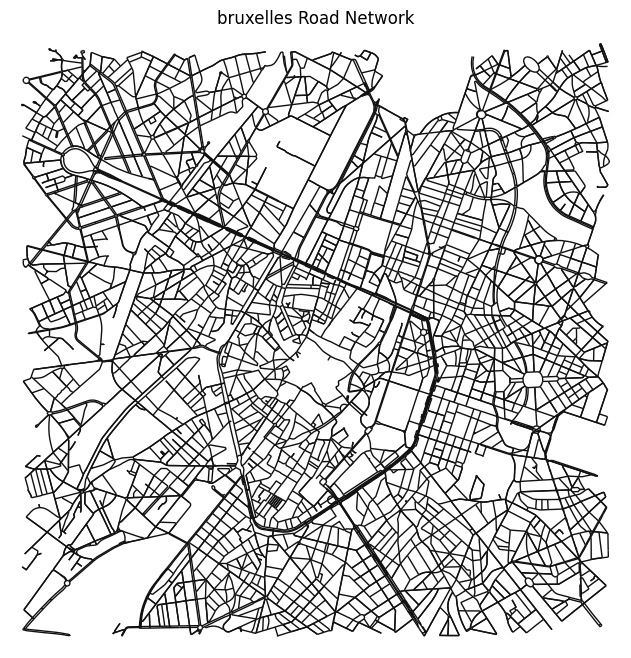}} \\
\hline
Buenos Aires & \grid{} & \adjustbox{valign=m}{\includegraphics[width=0.083\textwidth]{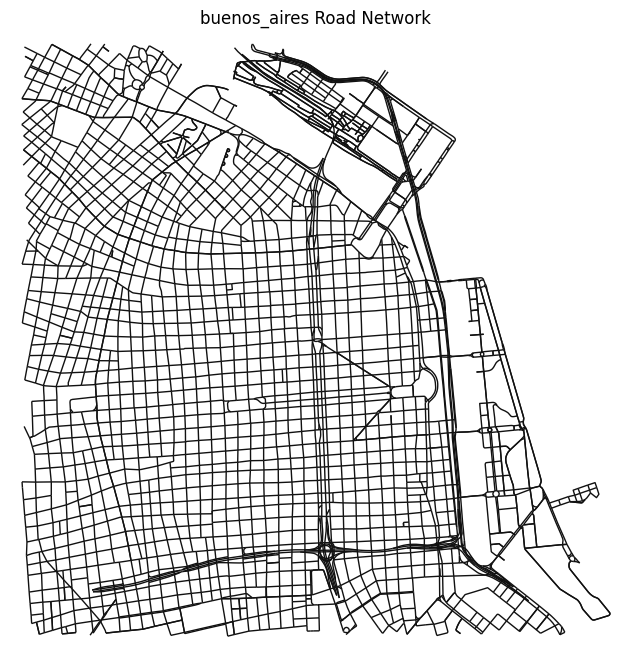}} & 
Cairo & \organic{} & \adjustbox{valign=m}{\includegraphics[width=0.083\textwidth]{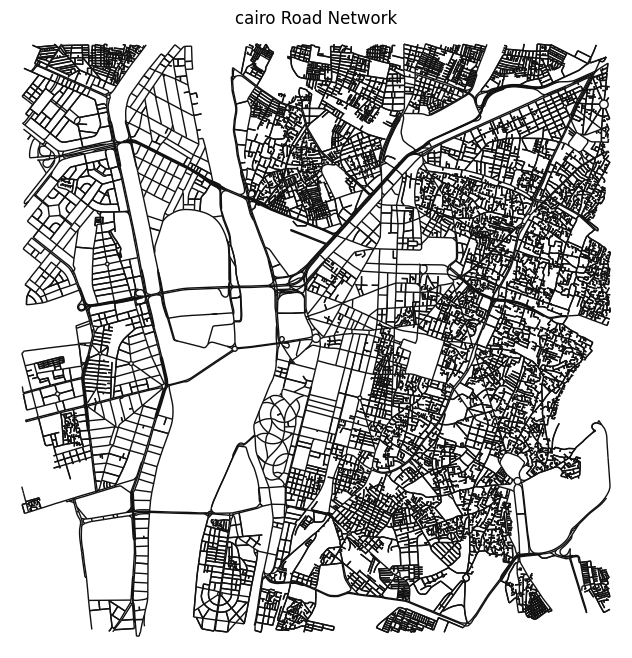}} & 
Cape Town & \organic{} & \adjustbox{valign=m}{\includegraphics[width=0.083\textwidth]{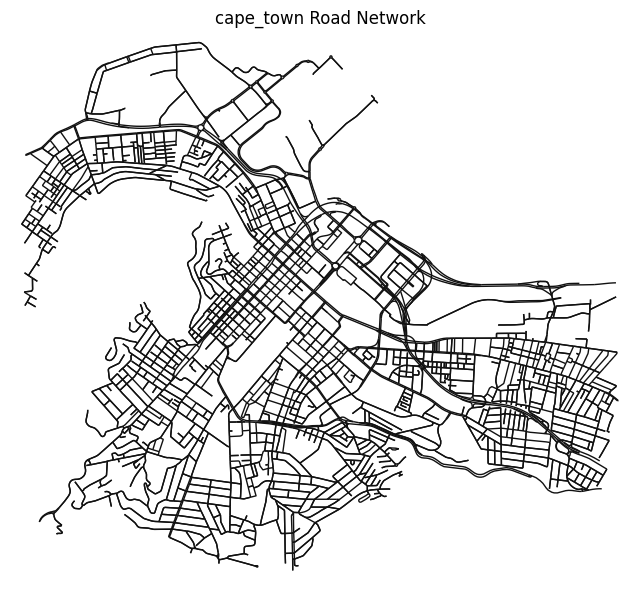}} \\
\hline
Dallas & \grid{} & \adjustbox{valign=m}{\includegraphics[width=0.083\textwidth]{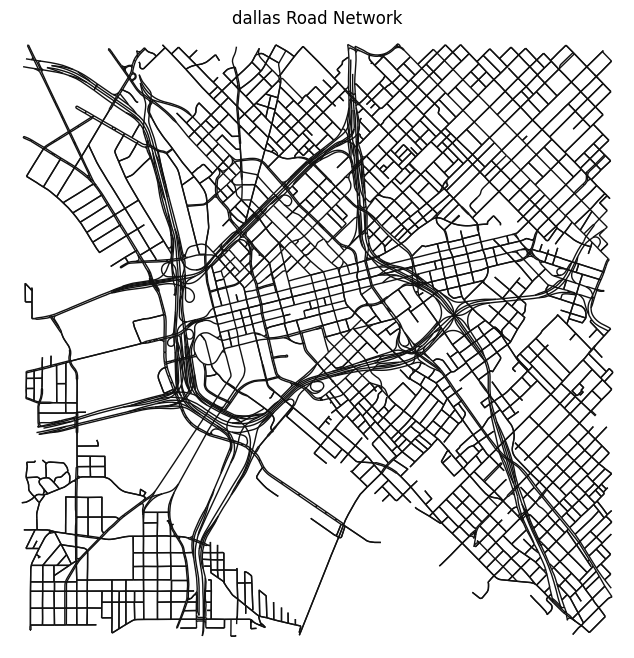}} & 
Dhaka & \organic{} & \adjustbox{valign=m}{\includegraphics[width=0.083\textwidth]{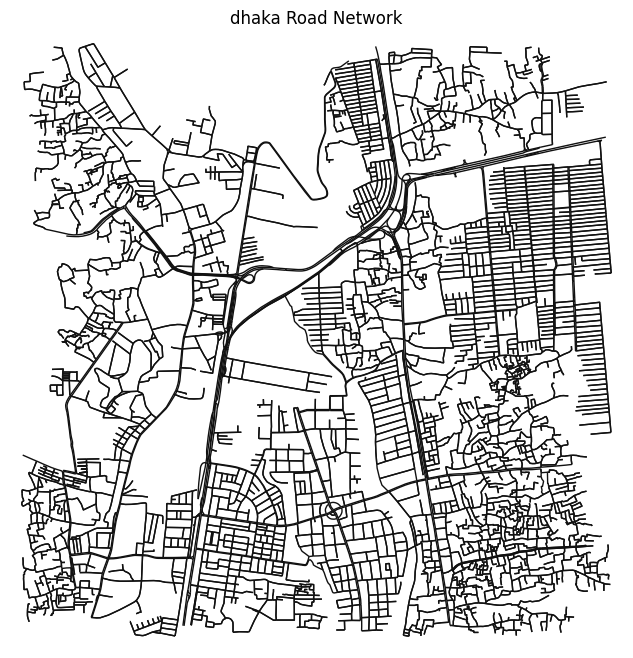}} & 
Dubai & \grid{} & \adjustbox{valign=m}{\includegraphics[width=0.083\textwidth]{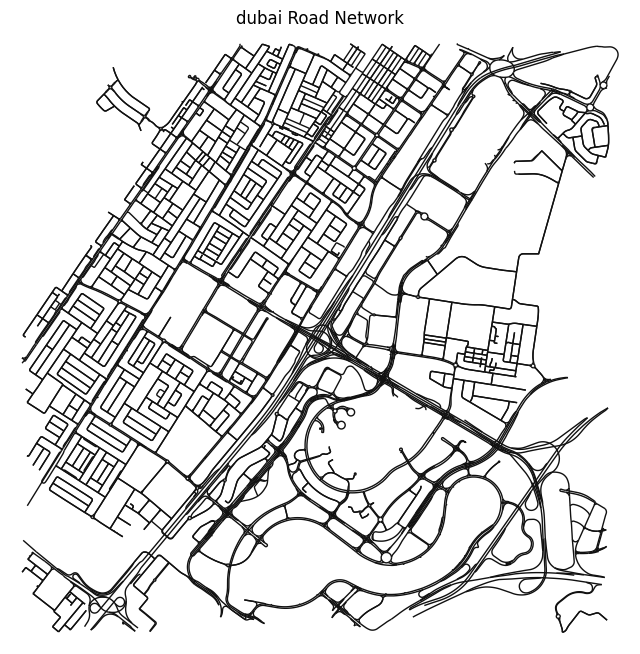}} \\
\hline
Guangzhou & \organic{} & \adjustbox{valign=m}{\includegraphics[width=0.083\textwidth]{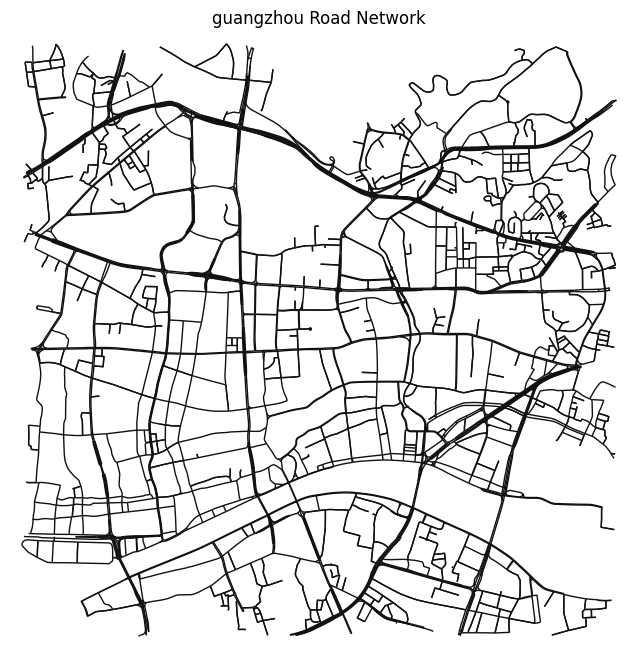}} & 
Hamburg & \organic{} & \adjustbox{valign=m}{\includegraphics[width=0.083\textwidth]{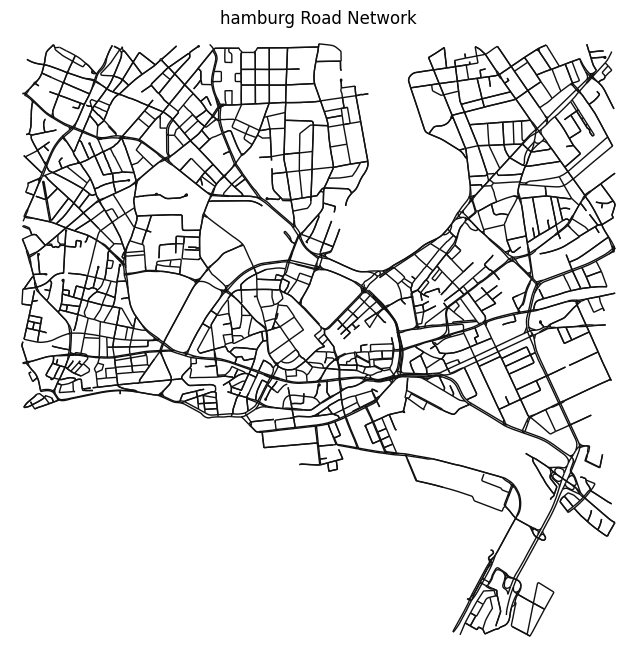}} & 
Istanbul & \organic{} & \adjustbox{valign=m}{\includegraphics[width=0.083\textwidth]{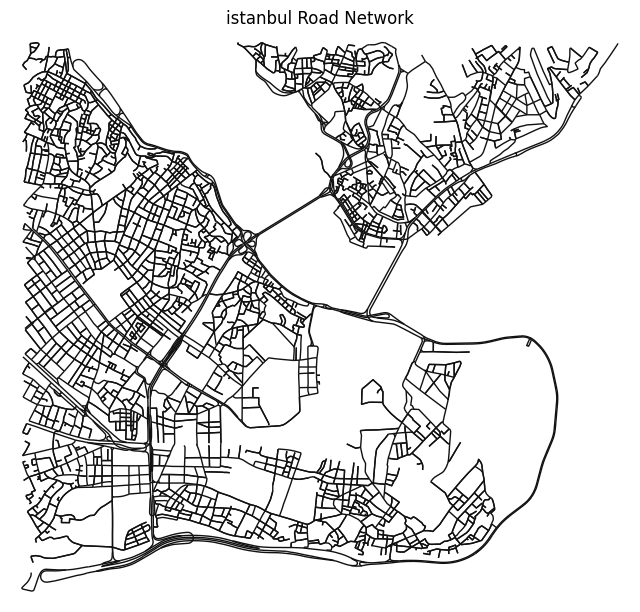}} \\
\hline
Jakarta & \organic{} & \adjustbox{valign=m}{\includegraphics[width=0.083\textwidth]{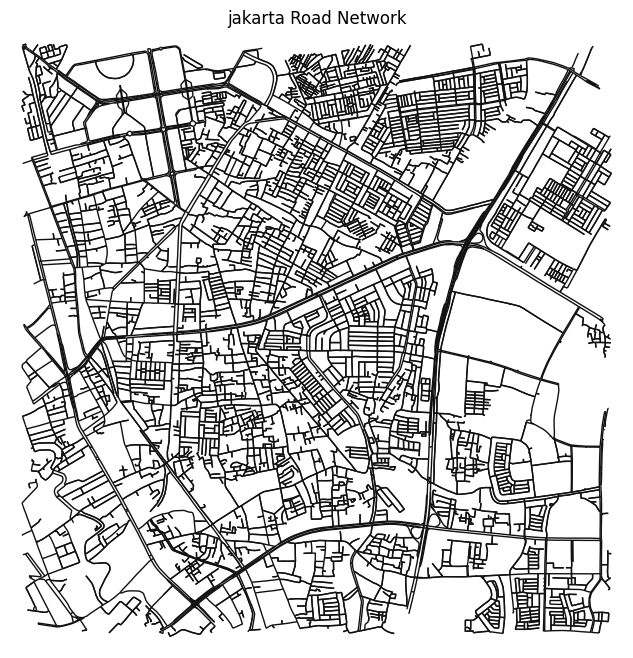}} & 
Karachi & \organic{} & \adjustbox{valign=m}{\includegraphics[width=0.083\textwidth]{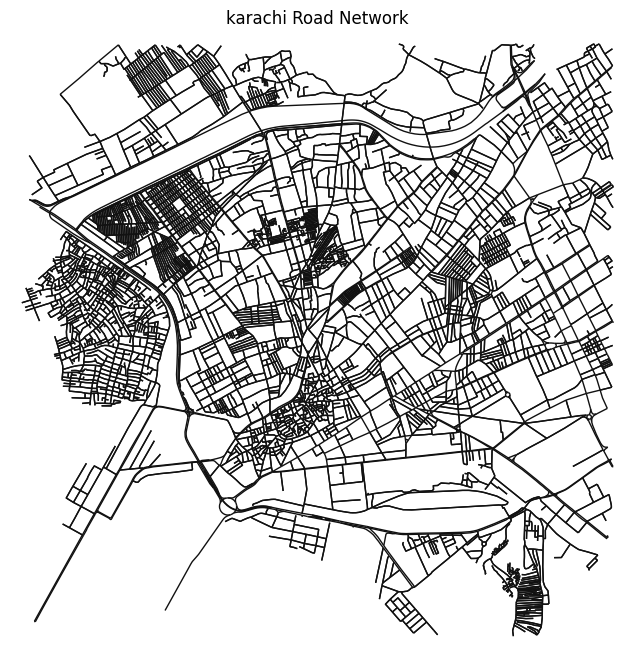}} & 
Kinshasa & \organic{} & \adjustbox{valign=m}{\includegraphics[width=0.083\textwidth]{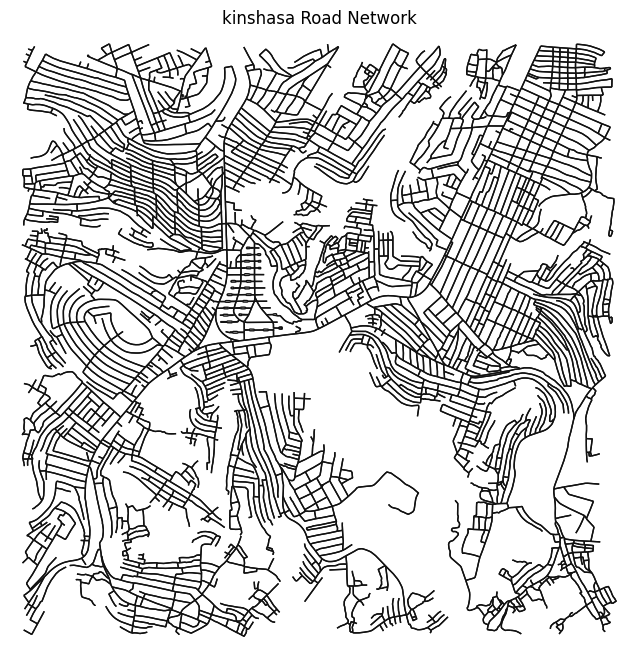}} \\
\hline
Lagos & \organic{} & \adjustbox{valign=m}{\includegraphics[width=0.083\textwidth]{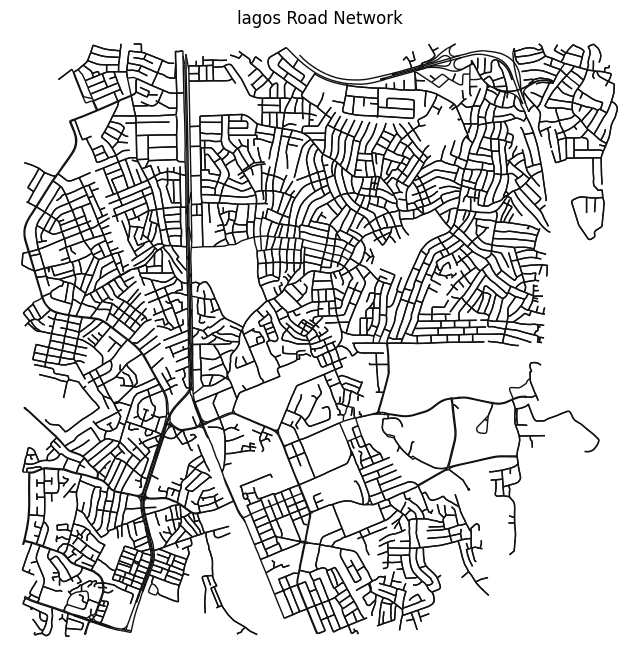}} & 
Lima & \grid{} & \adjustbox{valign=m}{\includegraphics[width=0.083\textwidth]{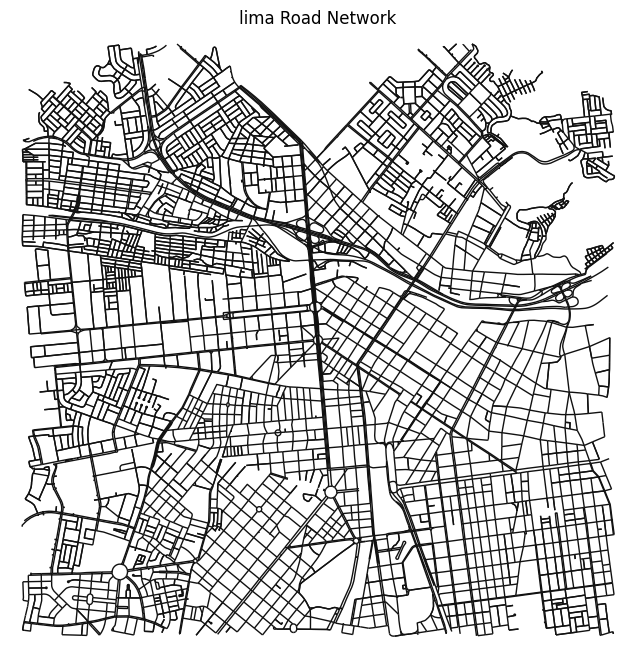}} & 
London & \organic{} & \adjustbox{valign=m}{\includegraphics[width=0.083\textwidth]{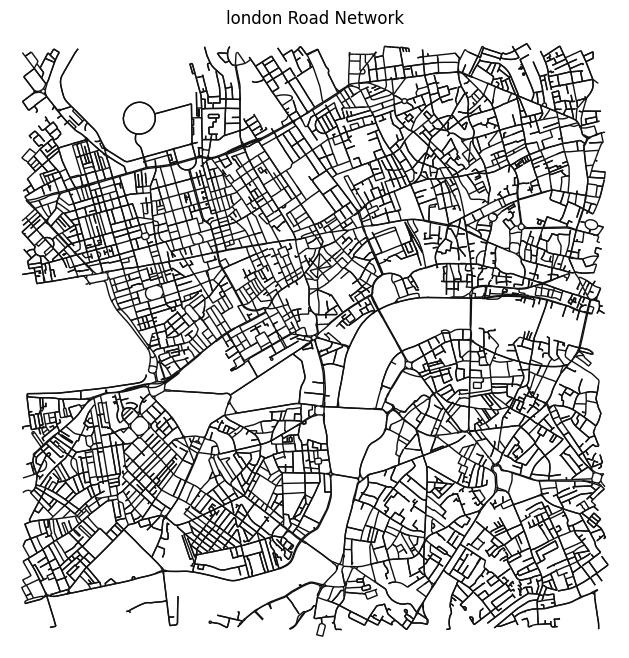}} \\
\hline
Los Angeles & \grid{} & \adjustbox{valign=m}{\includegraphics[width=0.083\textwidth]{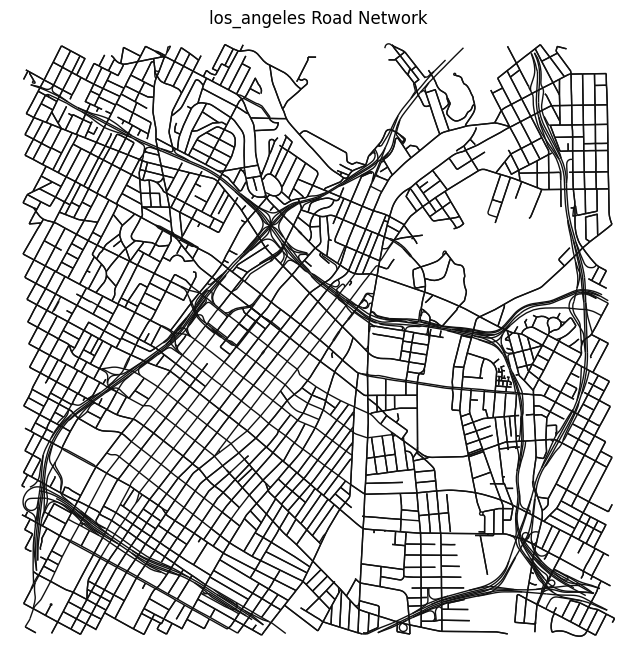}} & 
Manila & \grid{} & \adjustbox{valign=m}{\includegraphics[width=0.083\textwidth]{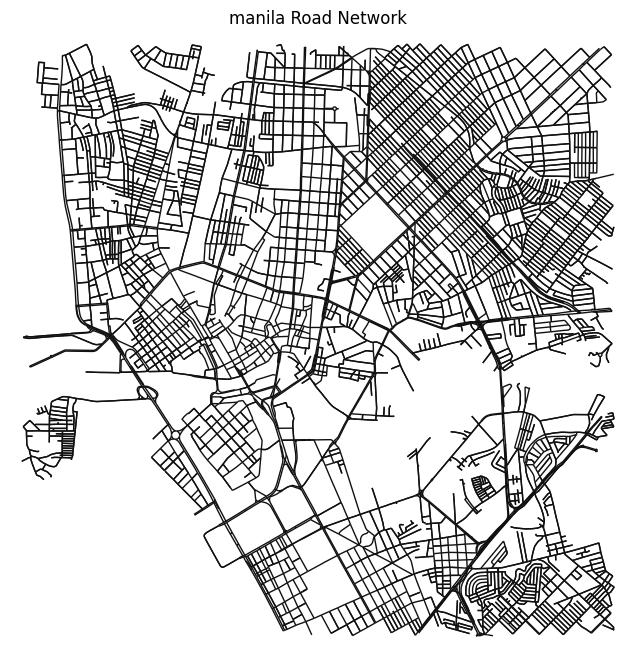}} & 
Melbourne & \grid{} & \adjustbox{valign=m}{\includegraphics[width=0.083\textwidth]{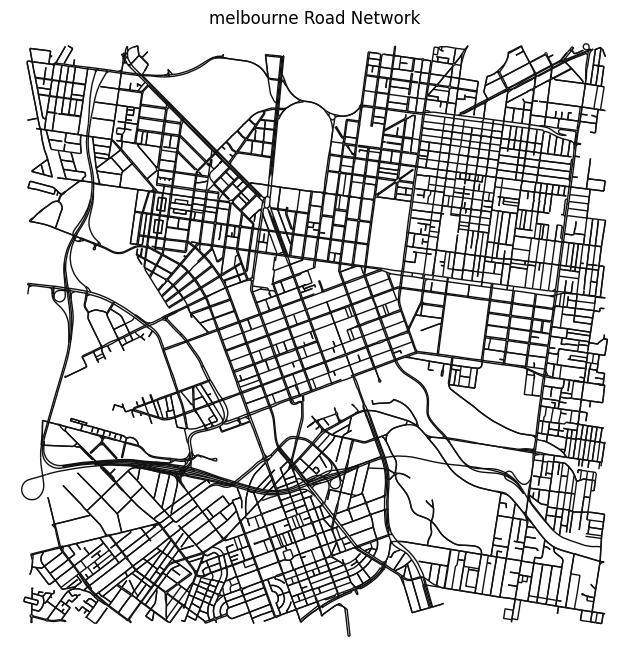}} \\
\hline
Mexico City & \grid{} & \adjustbox{valign=m}{\includegraphics[width=0.083\textwidth]{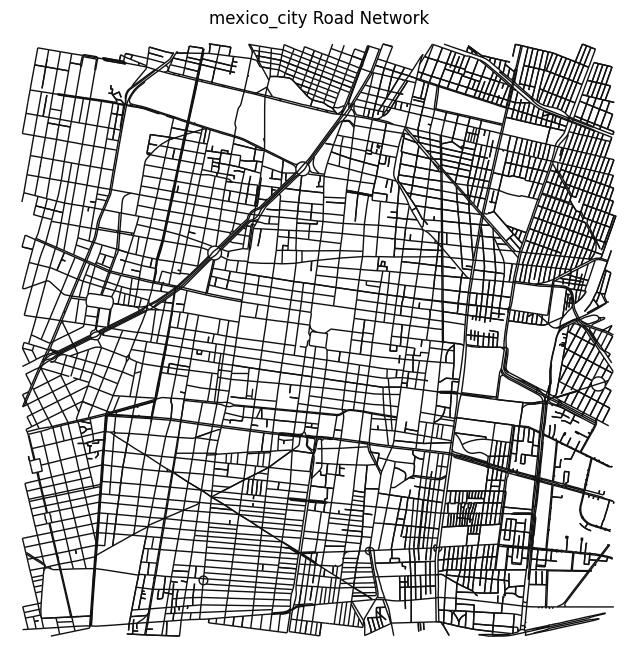}} & 
Milan & \radial{} & \adjustbox{valign=m}{\includegraphics[width=0.083\textwidth]{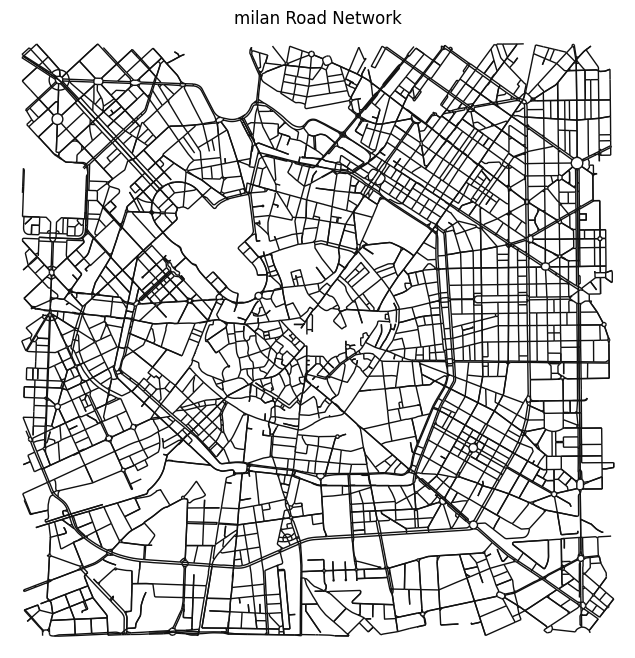}} & 
Moskow & \radial{} & \adjustbox{valign=m}{\includegraphics[width=0.083\textwidth]{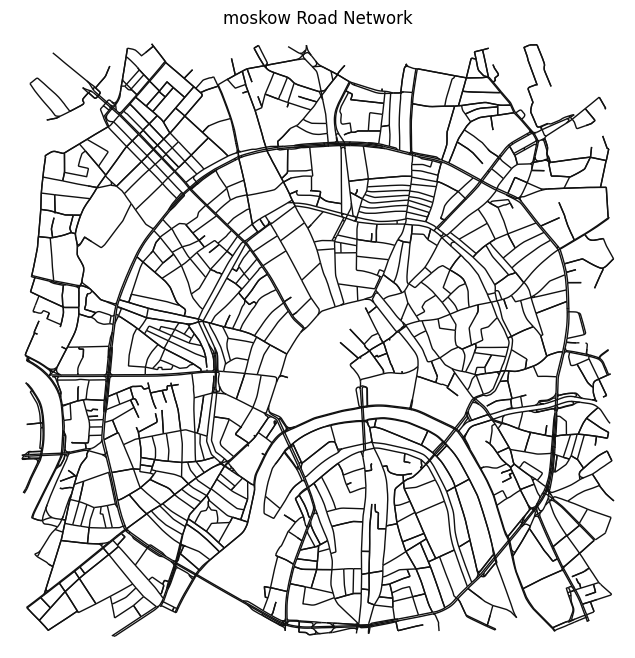}} \\
\hline
Mumbai & \organic{} & \adjustbox{valign=m}{\includegraphics[width=0.083\textwidth]{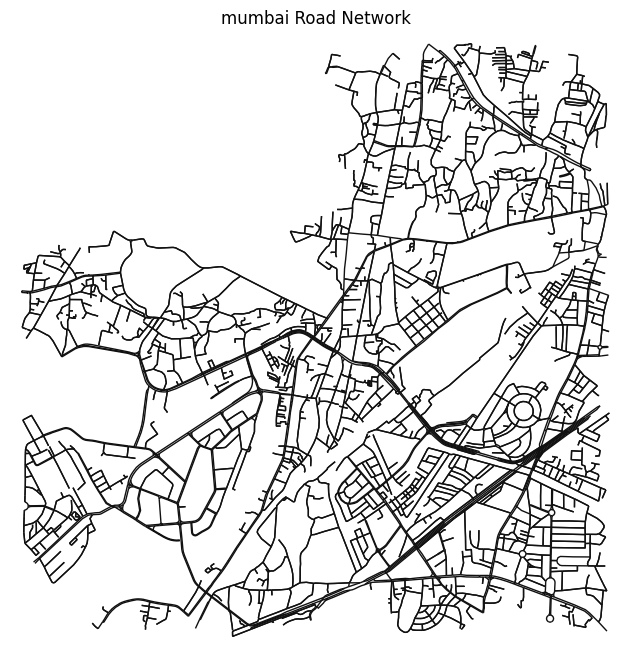}} & 
New Delhi & \organic{} & \adjustbox{valign=m}{\includegraphics[width=0.083\textwidth]{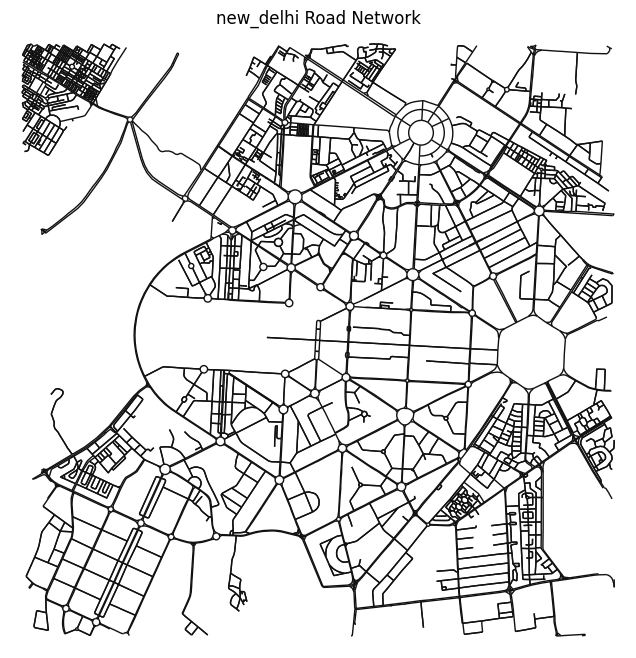}} & 
New York & \grid{} & \adjustbox{valign=m}{\includegraphics[width=0.083\textwidth]{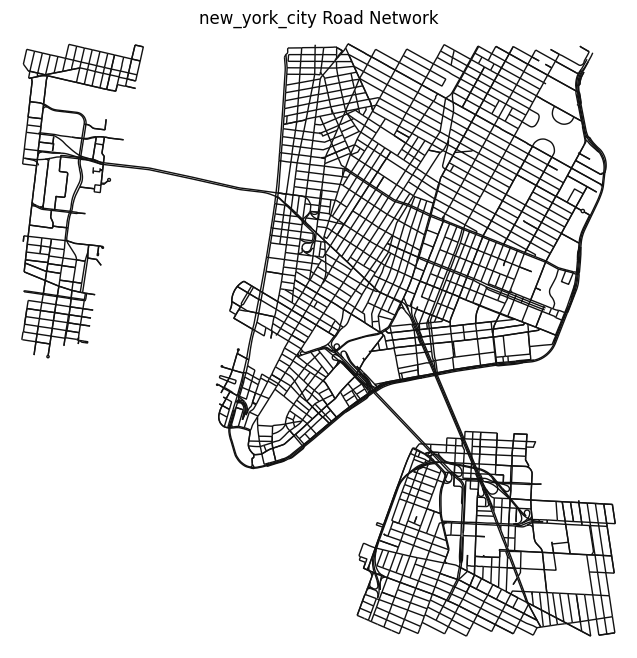}} \\
\hline
Osaka & \organic{} & \adjustbox{valign=m}{\includegraphics[width=0.083\textwidth]{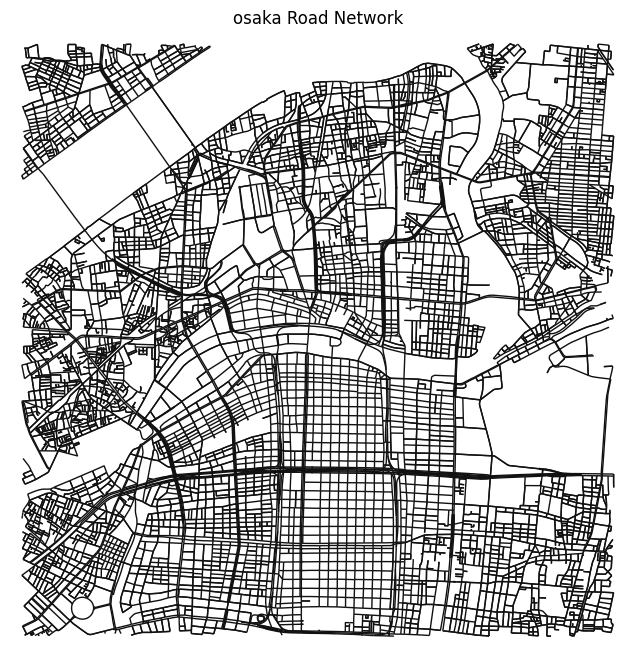}} &
Ottawa & \grid{} & \adjustbox{valign=m}{\includegraphics[width=0.083\textwidth]{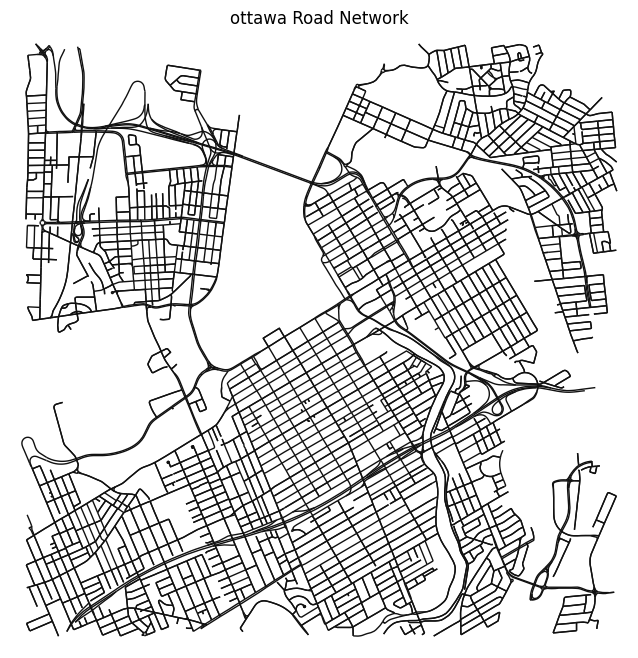}} & 
Paris & \organic{} & \adjustbox{valign=m}{\includegraphics[width=0.083\textwidth]{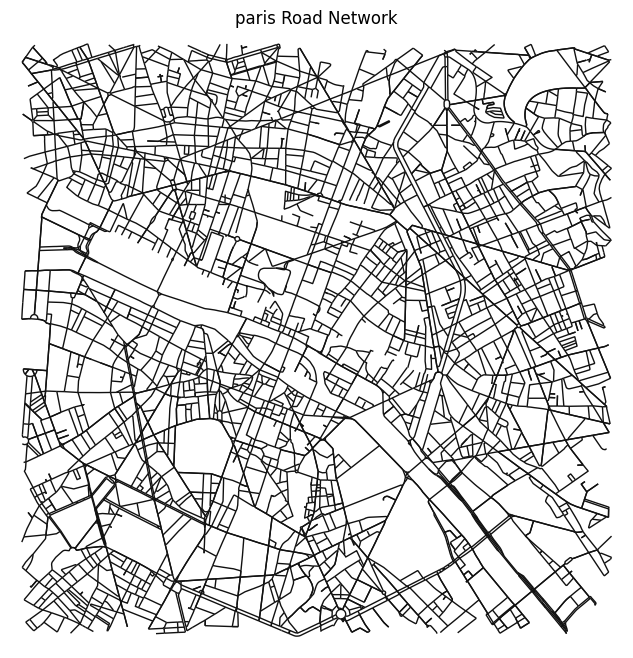}}  \\
\hline
Rome & \organic{} & \adjustbox{valign=m}{\includegraphics[width=0.083\textwidth]{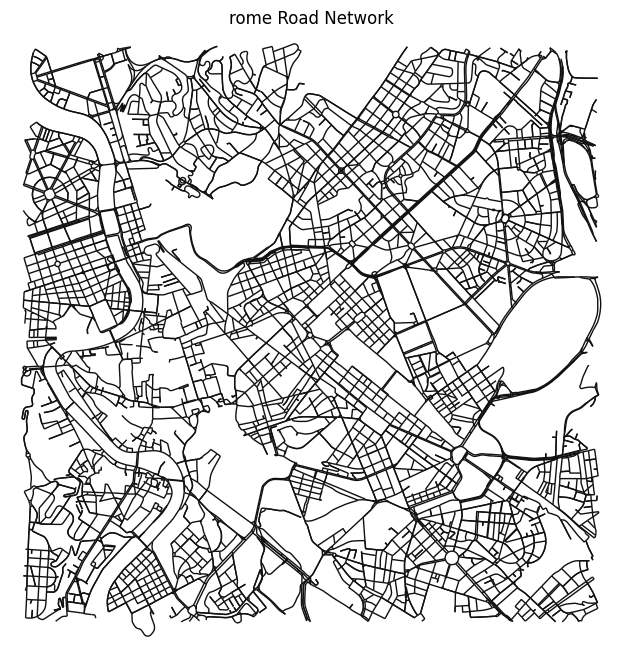}} &
Sao Paulo & \organic{} & \adjustbox{valign=m}{\includegraphics[width=0.083\textwidth]{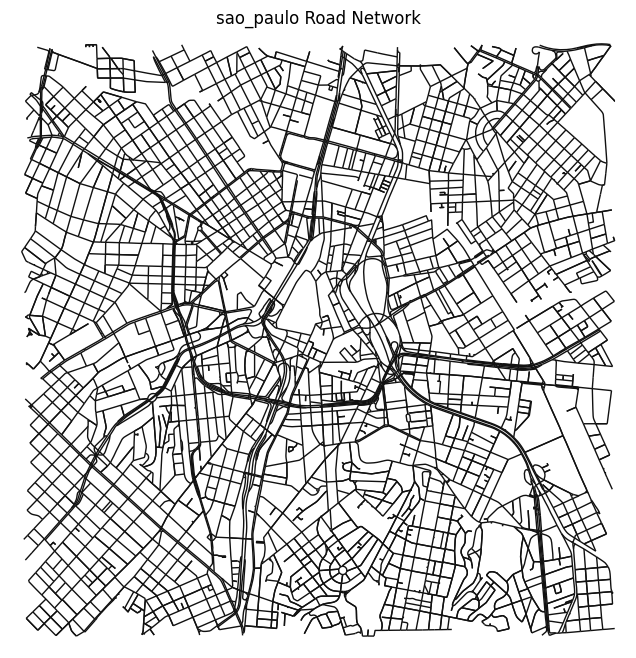}} & 
Seoul & \organic{} & \adjustbox{valign=m}{\includegraphics[width=0.083\textwidth]{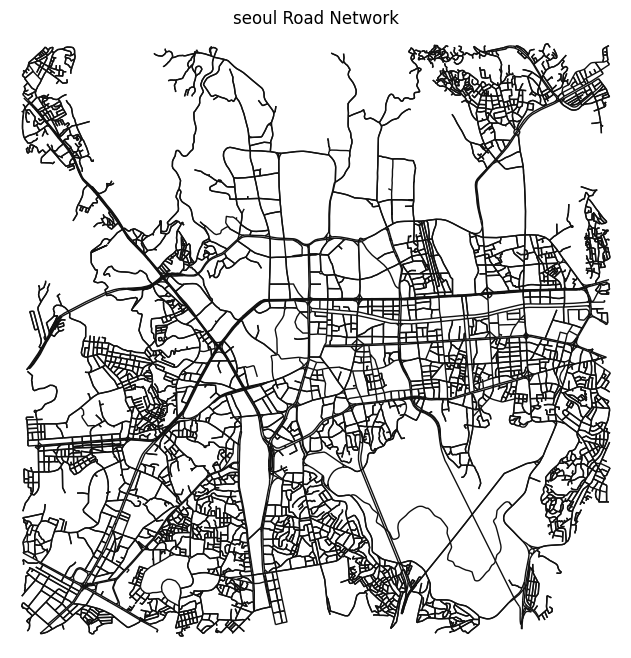}}  \\
\hline
Shangai & \organic{} & \adjustbox{valign=m}{\includegraphics[width=0.083\textwidth]{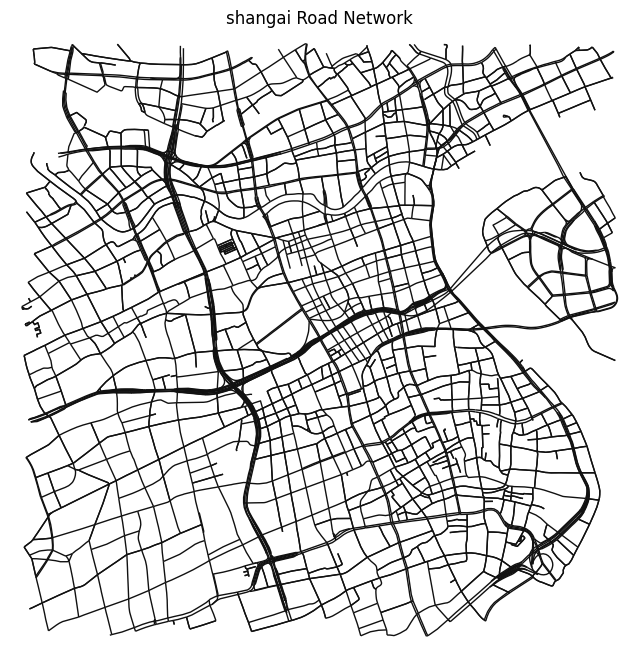}} &
Shenzen & \organic{} & \adjustbox{valign=m}{\includegraphics[width=0.083\textwidth]{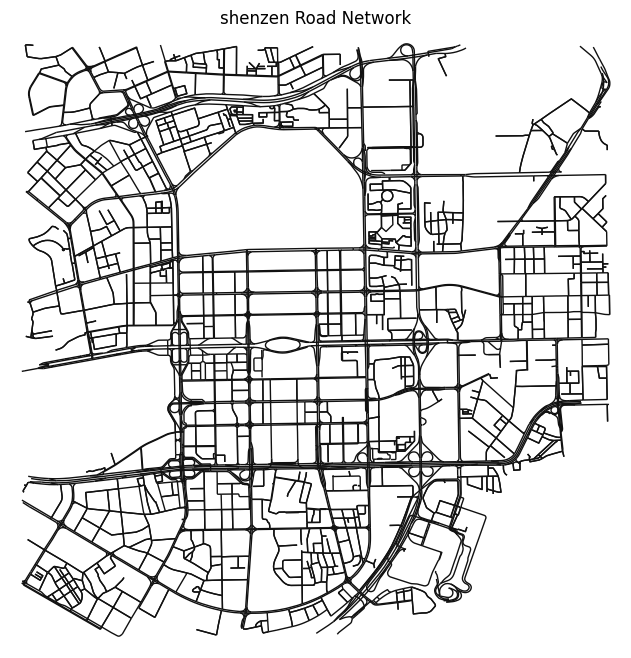}} & 
Sydney & \organic{} & \adjustbox{valign=m}{\includegraphics[width=0.083\textwidth]{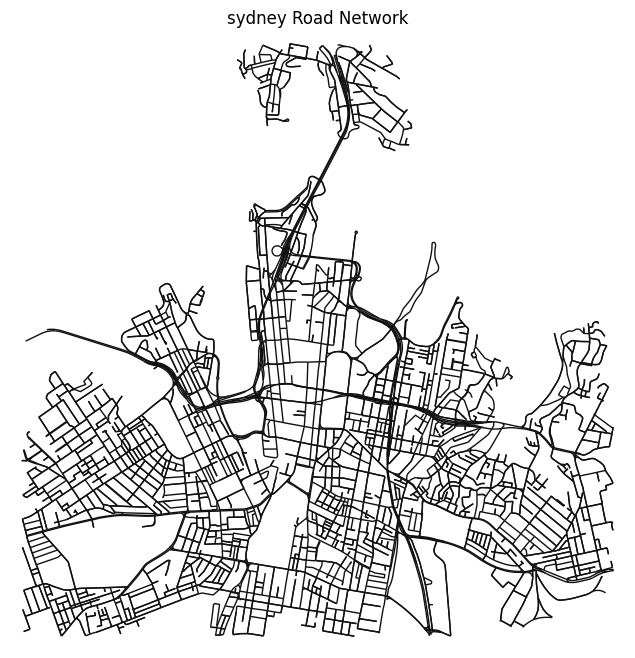}}  \\
\hline
Tehran & \organic{} & \adjustbox{valign=m}{\includegraphics[width=0.083\textwidth]{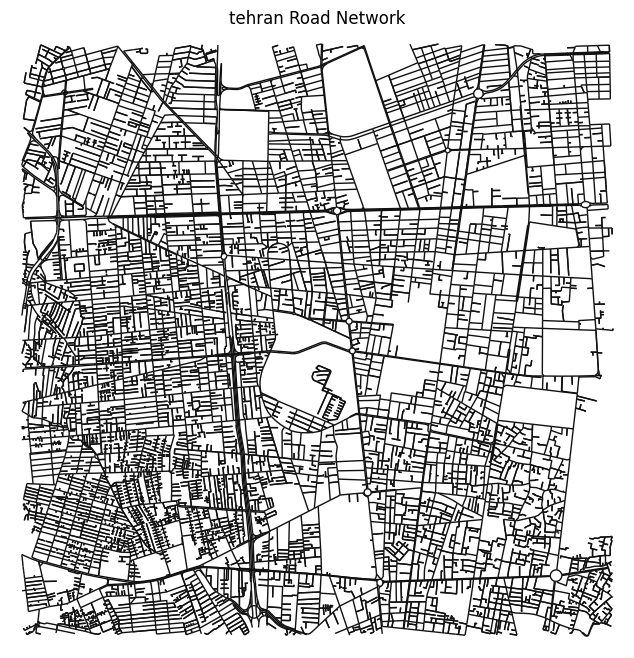}} &
Tokyo & \organic{} & \adjustbox{valign=m}{\includegraphics[width=0.083\textwidth]{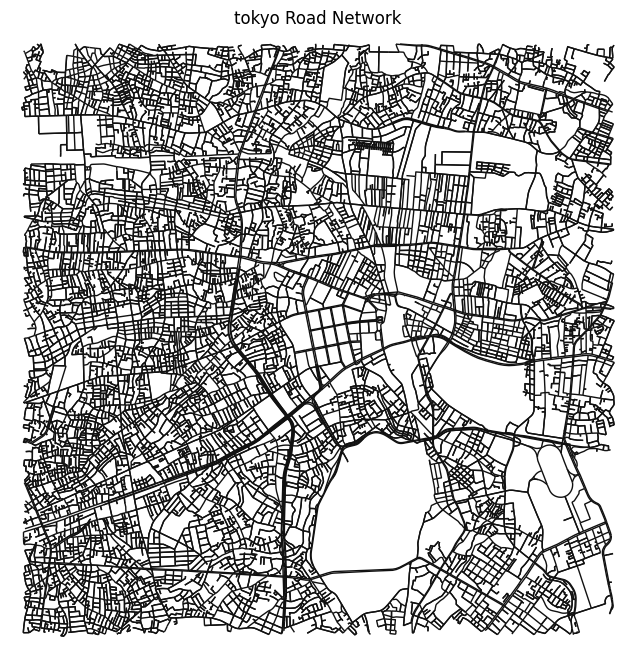}} & 
Toronto & \grid{} & \adjustbox{valign=m}{\includegraphics[width=0.083\textwidth]{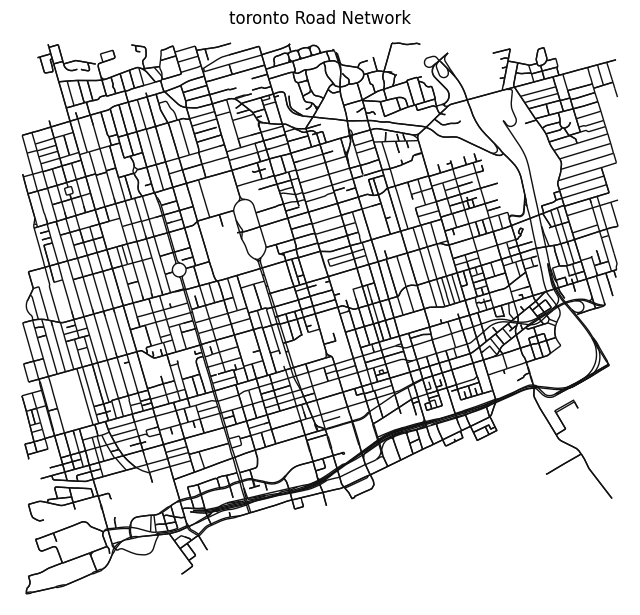}} \\
\hline
\multicolumn{6}{c}{} &
Vancouver & \grid{} & \adjustbox{valign=m}{\includegraphics[width=0.083\textwidth]{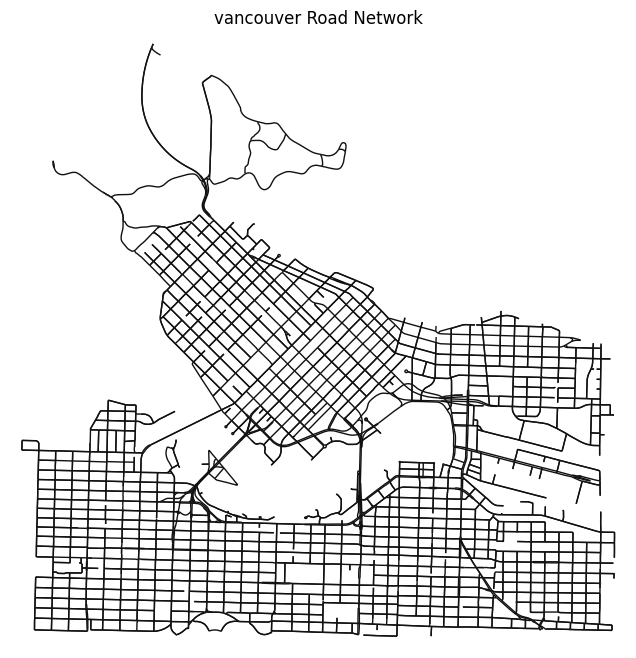}} \\[-0.5mm]
\cmidrule{7-9}
\end{tabular}
\vspace{-0.7cm}
\caption{City shapes and road network thumbnails.}
\label{tab:city_shapes}

\end{table}

Table~\ref{tab:stability_vs_shape} shows aggregate stability values grouped by road network shape, together with information on the length of path deviations, streets length and entropy of bearing. As we can see, the cities classified as Grid show a lower stability, followed by Organic cities and, as most stable category, Radial. 
These results generally confirm those obtained through entropy of bearing, and indeed the most unstable group of cities (Grid) is also the one with the lowest average entropy.
The Organic group arrives rather close to the Grid one along all the measures provided although the entropy of bearing is significantly larger. 
A closer inspection reveals that the two approaches (entropy vs. shape) simply model the concept of regularity in a different way. 
In particular, entropy enforces a rather strict requirement that recognizes as regular only quasi-perfectly grid structures (as in the case of the group of four very low-entropy cities highlighted in Figure~\ref{fig:stability_vs_entropy}, which indeed are all marked as Grid also in Table~\ref{tab:city_shapes}), whereas cities containing grid-like modules would be confounded with irregular ones, since different modules might be oriented in different directions, thus increasing the entropy of headings.
Similarly, Radial shapes -- though generally speaking more regular than Organic ones -- exhibit an even larger bearing entropy, showing that entropy is not a good criterion to identify them.
The study of urban shape, instead, is by definition more flexible, yet also (inevitably) slightly subjective, since human discernment is involved.

A comparison between the path deviations (in particular their normalized length, $R$) and the stability values show that Grid networks usually lead to longer paths to cover the perturbation introduced, thus suggesting that the lower stability of shortest paths might be related to the difficulty of reaching the new destination by just extending the path that reached the original one -- for instance because the new one ends up in a completely different street or because one-way roads make the path ``correction'' impossible.

\begin{table}[!h]
\resizebox{\textwidth}{!}{
\begin{tabular}{l| c|c c c c c}
\toprule
 & & & \textbf{length} & & \textbf{average} & \textbf{entropy} \\
\textbf{shape} & \textbf{n.} & $\mathbf{S_\Delta(o,d)}$ & $\mathbf{p(d, d^x)}$ \textbf{[m]} & \textbf{R} & \textbf{street length [m]} & \textbf{bearing} \\

\midrule
Grid & 15 &0.979 & 140.277 & 0.012 & 130.149 & 3.255 \\
\rowcolor{lavender!80} Organic & 27 & 0.984 & 122.864 & 0.010 & 119.206 & 3.511 \\
Radial & 4 &0.993 & 88.931 & 0.006 & 149.086 & 3.558 \\
\bottomrule
\end{tabular}}
\caption{Aggregate Shortest path stability $S_\Delta(o,d)$, length and normalized length (R) of path deviation $p(d,d^x)$, average street length and the entropy of street orientations vs. overall shape of the road network.}
\label{tab:stability_vs_shape}
\end{table}

\subsection*{Q6: Can cities be classified into homogeneous groups?}

We also investigate whether cities can be grouped into homogeneous clusters based on their routing characteristics and road network features, aiming to reveal shared structural attributes that influence path stability. This clustering approach highlights patterns across diverse urban layouts, creating stability profiles that reflect each cluster’s unique structural characteristics. It is important to note that stability was not used as a clustering attribute; rather, it is analyzed post-clustering to understand how stability relates to each group.
Our analysis considers several network and routing-related features, including the length of $p(d,d^x)$, the ratio between $p(d,d^x)$ and $p(o,d)$ ($R$), the average and standard deviation of street lengths, average road circuity, and entropy of street orientations. 
These values are reported in Table~\ref{tab:divercity_single_cities}.
Clustering was performed using the standard K-means clustering algorithm, determining an optimal number of clusters ($k=4$) through the elbow method~\cite{tan2007introduction}.

Each cluster, represented in Table~\ref{tab:cluster_metrics} by mean values of these attributes and overall stability, presents distinctive patterns of stability and network structure across cities. Cities in \textbf{Cluster 0} exhibit the highest route stability (0.993), likely due to shorter route length between original and perturbed destinations (89.89 meters) and lower ratio $R$. These cities have the longest and most varied street lengths, with an average of 172.65 meters and high variability, suggesting a complex layout with both long and short segments. The relatively high road circuity (1.08) indicates winding roads, and the high entropy of street orientations, together with a slight predominance of Organic cities suggests a non-regular road network structure.

\textbf{Cluster 1} is the largest group, including 24 cities with moderately high stability (0.982) and shorter average street lengths. The road circuity is lower than in Cluster 0 (1.06), indicating straighter and more direct roads. These cities are mainly characterized by non-grid layouts, as reflected by their high entropy of street orientations (3.9) and the strong presence of Organic shapes. They also have shorter and less varied street lengths, along with a low $R$, further supporting stability, as destinations and their displacements are well-connected.

\textbf{Cluster 2} exhibits lower stability (0.977) and is associated with the lowest entropy of street orientations (2.83), typical of cities with a regular, grid-like road network -- and indeed it contains only Grid shapes. Although their length of route $p(d,d^x)$, circuity, and the ratio $R$ are similar to those in Cluster 1, the grid-like structure contributes to slightly reduce stability.

\textbf{Cluster 3} has the lowest stability (0.971) due to the highest ratio $R$ and the greatest deviation distance between original and perturbed destinations, which forces substantial detours resulting in high path divergence. This cluster also exhibits the lowest circuity (1.05), suggesting a network with primarily straight roads that lack the flexibility seen in clusters with higher circuity. The cities in this cluster typically feature short and less varied street lengths, containing a mix of Organic and Grid road networks -- the latter probably representing borderline cases of their shape category, since the average entropy of orientations if rather high (3.47). 

\medskip

In summary, this clustering analysis identifies key factors for stable and unstable path profiles within urban road networks. Stable profiles are associated with cities that have shorter route deviations (length of $p(d, d^x)$), low $R$, longer, varied street segments, and higher orientation entropy (and indeed, they contain a smaller proportion of Grid shapes). In contrast, unstable profiles are found in cities where destination perturbations result in higher deviation distances and greater $R$ values, forcing substantial detours. Unstable path profiles are often linked to cities with low circuity, short, uniform street lengths, and lower orientation entropy, mainly associated to Grid layouts.

\begin{table}[!h]
\resizebox{\textwidth}{!}{
\begin{tabular}{l|c|c c c c c c|c}
\toprule
 &  & \textbf{length} &  & \textbf{avg. street} & \textbf{std street} & \textbf{average} & \textbf{entropy} & \\
\textbf{Cluster} & $\mathbf{S_\Delta(o,d)}$ & $\mathbf{p(d, d^x)}$ \textbf{[m]} & \textbf{R} & \textbf{length [m]} & \textbf{length [m]} & \textbf{circuity} & \textbf{of bearing} & \textbf{G-R-O} \\\midrule
0 & 0.993 & 89.889 & 0.006 & 172.646 & 255.944 & 1.076 & 3.494 & 2-3-7 \\
\rowcolor{lavender!80} 1 & 0.982 & 125.693 & 0.010 & 103.908 & 135.172 & 1.058 & 3.492 & 5-1-18 \\
2 & 0.977 & 124.384 & 0.012 & 150.942 & 186.241 & 1.072 & 2.826 & 4-0-0 \\
\rowcolor{lavender!80} 3 & 0.971 & 197.395 & 0.019 & 99.640 & 146.042 & 1.053 & 3.469 & 4-0-2 \\
\bottomrule
\end{tabular}}
\caption{Output clusters obtained by K-means over city indicators, including the number of cities in the different shape categories (Grid-Radial-Organic). Notice that \textit{stability} and shape information are only used as explanatory variables, not involved in the clustering.}
\label{tab:cluster_metrics}
\end{table}

\subsection*{Q7: Are there spatial patterns of instability?}
\begin{figure}
    \centering
    \includegraphics[width=0.44\linewidth]{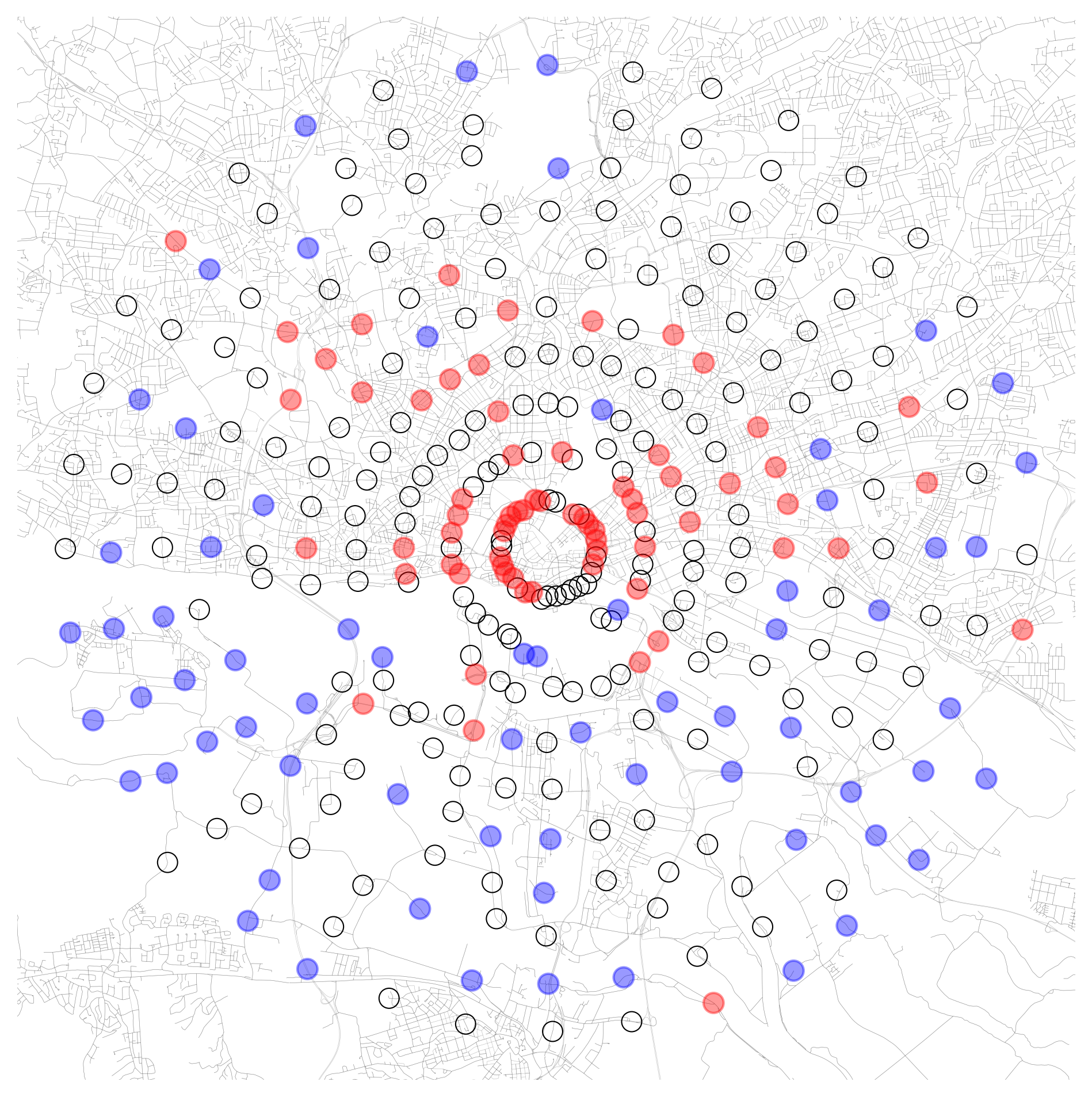} 
    \includegraphics[width=0.44\linewidth]{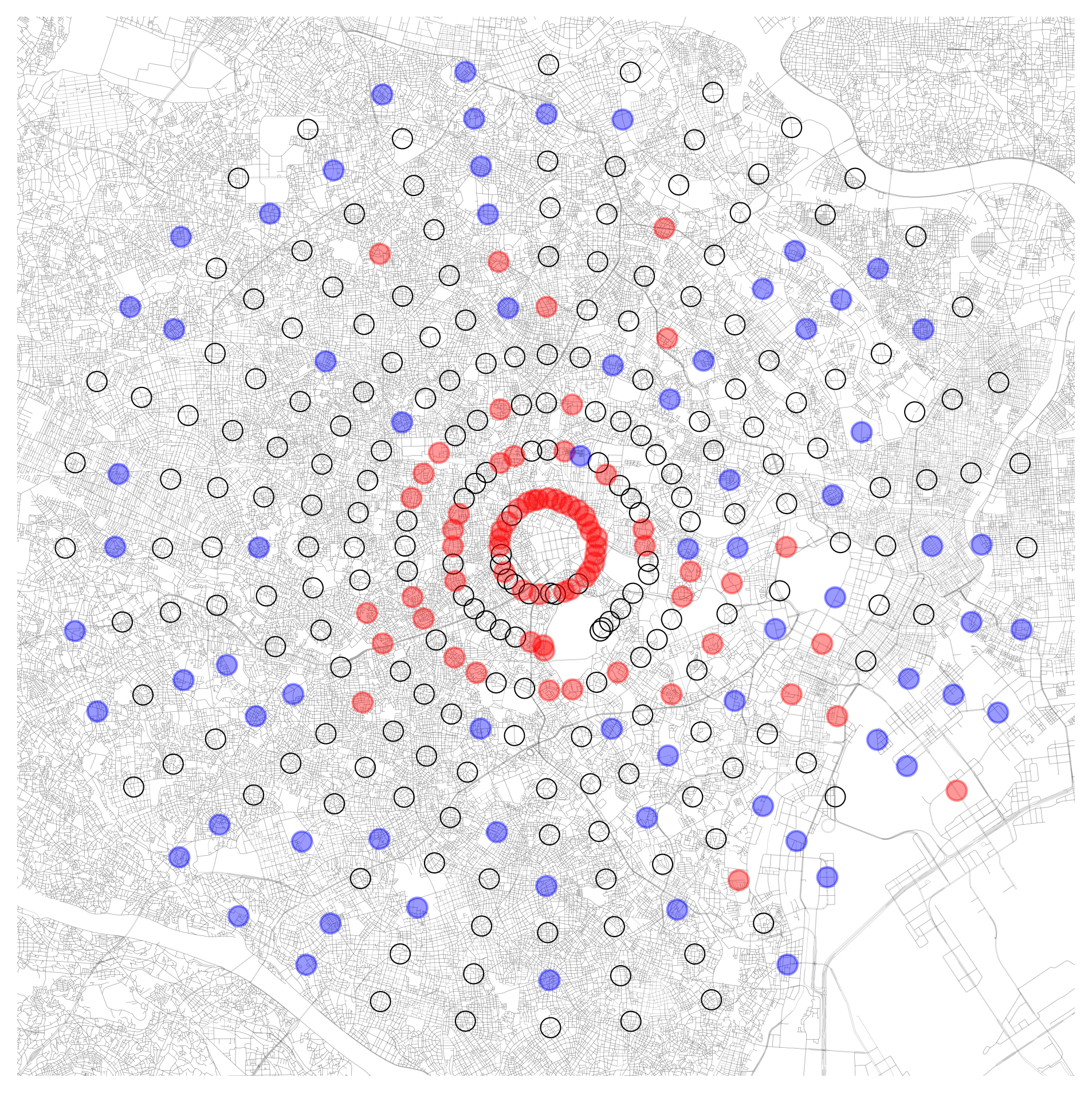} \\
    \texttt{Cluster 0: Hamburg \hspace{2cm} Cluster 1: Tokyo } \\
    \includegraphics[width=0.44\linewidth]{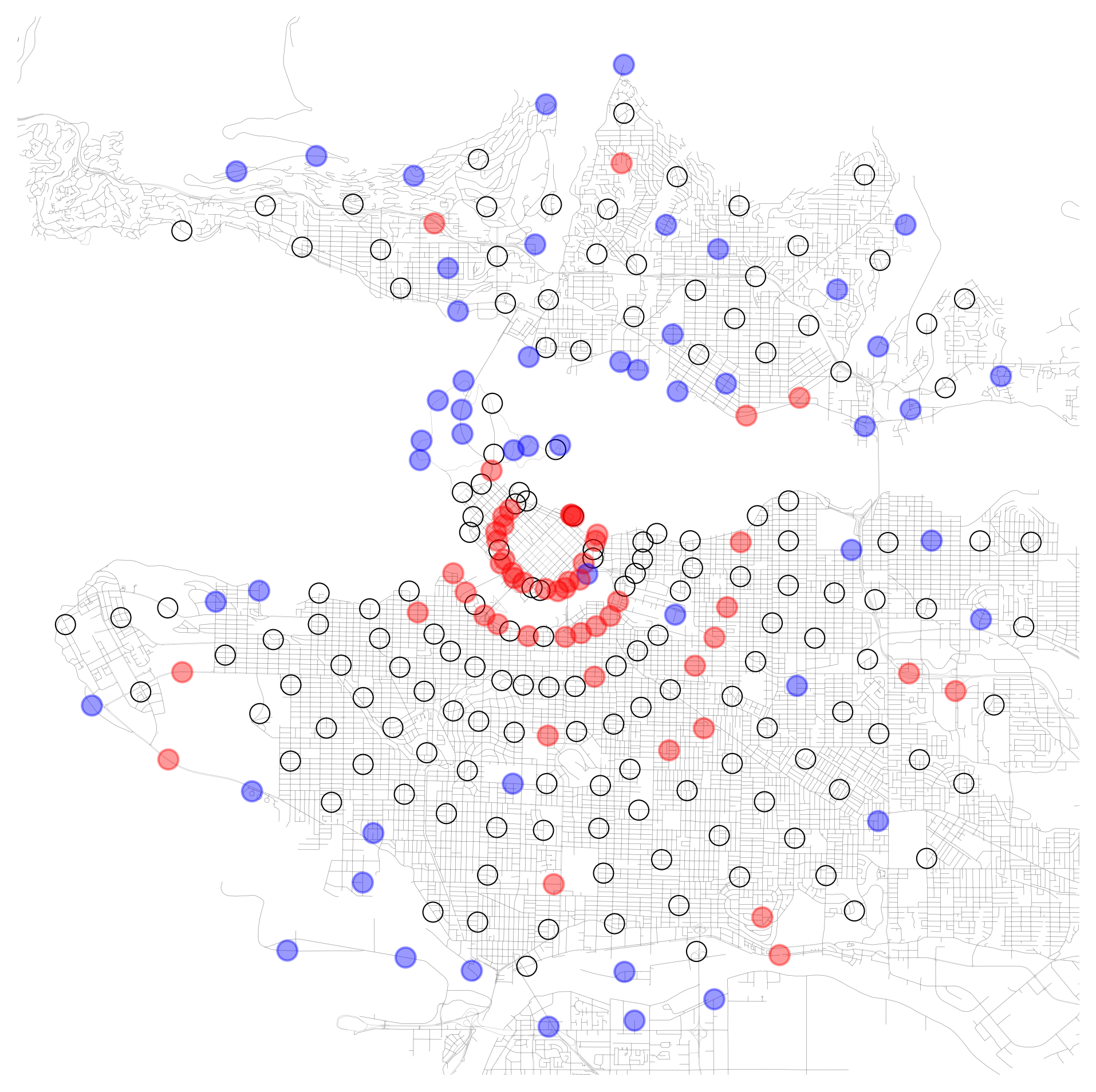} 
    \includegraphics[width=0.40\linewidth]{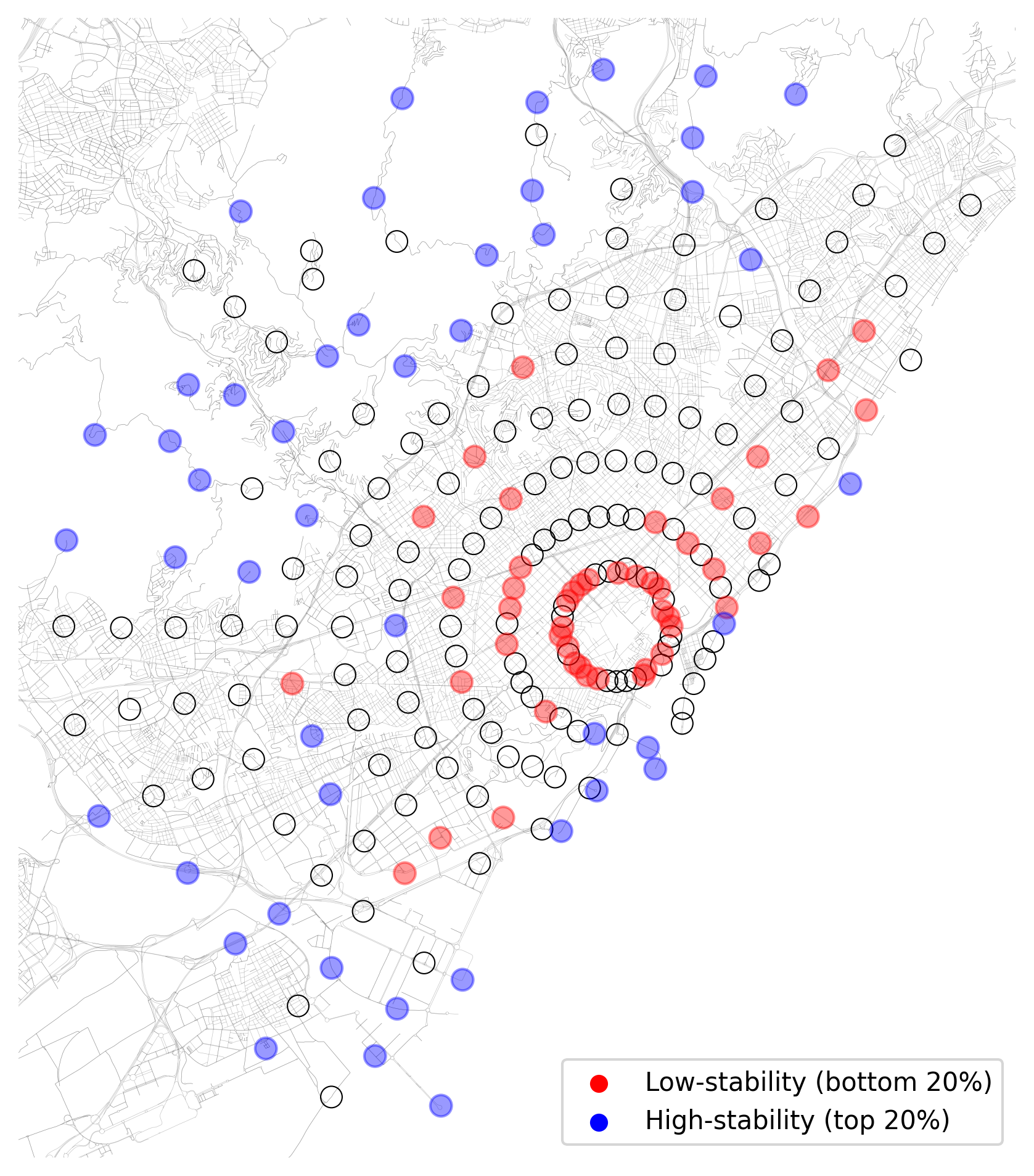} \\
    \texttt{\hspace{1pt} Cluster 2: Vancouver \hspace{2cm} Cluster 3: Barcelona}

    \caption{Spatial distribution of stable (blue) and unstable (red) destinations in four cities.}
    \label{fig:spatial_distribution}
\end{figure}

In order to answer to this question, we first define the stability $S_\Delta(d)$ of a destination location $d$ as the average of the stability values obtained considering all possible origins under the sampling described in Section~\ref{sec:pairs_gen}, namely $S_\Delta(d) = avg_o S_\Delta(o,d)$.
Then, we study the geographical distribution of $S_\Delta(d)$.
We do that through visualization over four highly diversified sample cities, each belonging to a different cluster of the clustering discussed above: Hamburg (Cluster 0), Tokyo (Cluster 1), Vancouver (Cluster 2) and Barcelona (Cluster 3). 
The results are shown in Figure~\ref{fig:spatial_distribution} where destinations up to 10 km from the city center are considered and, for the sake of readability, locations are colored based on their stability value: values below the 20th percentile are shown in red (unstable locations); those above the 80th percentile are in blue (stable locations); the others are depicted as empty circles.

We can observe a few common patterns. 
First, unstable destinations are mostly concentrated in the central areas, whereas stable ones cover higher distances from the center. This confirms the general trend shown in Figure~\ref{fig:stability_vs_radius}. 
Second, stable destinations are more concentrated in areas with a sparse local road network, as it happens in the North-West of Barcelona and the South of Hamburg. Roads in Tokyo are dense everywhere, and indeed the stable destinations are distributed more homogeneously.
A special case is Vancouver, where the harbor divides the city in two areas, connected by two bridges: this highly constrains paths connecting origins in one part to destinations in the other, reducing the availability of choices -- and thus increasing stability, especially of the Northern part, which is smaller and then most of its OD pairs start from the other side and cross the bridge.
Third, there is a clear spatial autocorrelation, namely unstable destinations tend to cluster together, rather than disperse randomly, not only in the center (where it is straightforward for their high density) but also in farther layers of the city.

\section{Discussion}
\label{sec:discussion}

\textbf{A universal pattern.}
The phenomenon of shortest path instability is not limited to our motivating example of Barcelona, which indeed exhibits high instability levels confirming our initial intuition, but instead manifests universally, albeit to varying extents. A consistent trend emerges: as distance from the city center increases, route stability grows and eventually plateaus in peripheral areas. While this stability trend is universal, each city maintains its own baseline level and rate of stability increase, influenced by distinct urban layouts, connectivity, and geographic factors.
\medskip

\textbf{Why does distance from the city center impact stability?}
Figure \ref{fig:stability_vs_radius} shows that path stability increases with distance from the city center, following an exponential trend that levels off at near-complete stability beyond 10 km. Low stability in the city center (0-5 km) likely results from dense and complex road networks characterized by frequent intersections. Furthermore, short OD distances in these areas mean that $p(d, d^x)$ represents a larger fraction of the original path $p(o,d)$, contributing to lower stability, as demonstrated by the negative relationship between $R$ and stability. Beyond 10 km, trips exhibit greater stability, likely due to the transition into road networks that feature long-range, fast routes (such as highways, motorways, etc.). Trips gravitating at this distance from the city center tend to flow into such long-range connections, that represent the main part of the trip and remain unchanged when slightly perturbing the destination.
\medskip

\textbf{The role of the local context on perturbed destinations and stability.}
The relationship between the original destination $d$ and its perturbed counterpart $d^x$ plays a critical role in determining path stability. If the route between $d$ and $d^x$ is lengthy or indirect compared to their displacement (i.e. the straight distance between $d$ and $d^x$), this suggests poor connectivity between $d$ and $d^x$, potentially due to geographic or network constraints. In such cases, the shortest route $p(o,d)$ is less likely to align with the shortest route $p(o,d^x)$, as weak connectivity reduces the chance of overlapping shortest paths. Instead, totally different alternative routes from $o$ to $d^x$ might be more efficient compared to simply extending the original route $p(o,d)$ with an extra leg $p(d,d^x)$ to reach $d^x$. This indicates that path instability is strongly driven by connectivity limitations within the network. Conversely, if $d$ and $d^x$ are well-connected, allowing for a short and direct route, any divergence between routes $p(o,d)$ and $p(o,d^x)$ likely reflects the network's adaptability and the presence of multiple efficient paths rather than structural or connectivity constraints.
\medskip

\textbf{Not only the perturbed destination matters. The origin does too.}
The perturbed destination is essential to path stability, but the position of the origin, particularly the distance separating the origin from the destination, is equally pivotal. When the length of the original route from $o$ to $d$ is short, even a moderate perturbation cost can result in completely disjoint paths, as reaching the perturbed destination $d^x$ may require a new directional approach rather than a simple extension of the original route.
In such cases, characterized by an high value of the ratio $R$, it is often more efficient for the shortest path algorithm to determine an entirely different route to $d^x$, resulting in lower stability. 
However, as the length of the original route $p(o, d)$ increases, $R$ decreases and the influence of the perturbation diminishes. 

\medskip

\textbf{The role of street length.}
As highlighted in Figure\ref{fig:stability_vs_elen}a, longer streets tend to be associated with higher stability. This may happen as higher average street length may imply fewer intersections or ``decision points'' where alternative paths can diverge. In cities with shorter street segments, paths can branch more frequently, increasing the likelihood of different routes being chosen when a destination is slightly perturbed. This could lead to lower stability as multiple potential shortest paths exist. Conversely, in cities with longer streets, fewer choices at intersections may limit route variability, resulting in greater stability.
\medskip

\textbf{City path stability: a road network fault or a feature?}

At the end of our journey, the natural question that arises is about the connotation we should give to stability. Is it a symptom of an issue, or an evidence of a virtuous city planning?
While a definitive answer is far from our current reach, we can shed some light by recalling some of the insights provided along this paper.
We do that observing that in terms of the role of the origin and destination in path stability there are primarily two scenarios driven by road network structure and connectivity. 

First, in dense, flexible networks, even small displacements in the destination can prompt alternative routes due to abundant intersections and route choices, where the shortest path algorithm might select a new path rather than simply extending the original one. This type of instability reflects the network’s adaptability, as alternative routes emerge naturally from minor destination shifts, suggesting a resilient layout and a virtuous road infrastructure plan. 

In contrast, a low stability can also be ``enforced'' when the perturbed destination is poorly connected to the original, as weak connectivity between $d$ and $d^x$ compels the algorithm to determine an entirely different path. Here, path divergence stems not from choice but from the network’s structural limitations, where insufficient local connectivity (low density of roads, inefficient one-way streets, etc.) prevents a seamless transition between nearby points. 

In summary, path stability appears to be by itself a neutral characteristic of a road network, and instability can either reflect network flexibility with abundant route options or structural limits/constraints that mandate an entirely new route.

In more applicative terms, knowing the stability of an area can help in evaluation and planning tasks.
Low stability indicates the presence of diverse alternative routes, which can be advantageous for dispersing traffic, especially in high-density events. For example, if an arena’s surroundings exhibit low stability, slight shifts in destination could distribute incoming traffic over multiple routes, easing congestion near the event site.
However, in settings where planners have designed high-capacity highways to manage and streamline large volumes of traffic, lower stability might counter these objectives. Stability around highways, where primary, high-efficiency routes are preferred, ensures that traffic is concentrated along these main arteries rather than dispersed across alternative paths. Thus, low stability near highways is not inherently negative; rather, it reflects a network structure most likely intentionally designed to focus traffic flow through key conduits.

\section{Conclusion}
\label{sec:conclusion}

In this work, we examine the stability of the shortest path in response to small displacements of the destination location. 
We analyze path stability across 46 major cities worldwide, focusing on how network structure, local context, and spatial patterns affect path stability.
Our findings reveal that path stability is a universal phenomenon, but it varies significantly with the road network's properties and trip characteristics. 
In particular, we highlighted significant relations between path stability and the trip's position (center vs. periphery), street lengths, the shape of the road network, length of the trips and reachability of the perturbed destinations.
In conclusion, path stability provides a valuable lens for evaluating the adaptability and resilience of urban transportation networks. High stability offers predictability, which is particularly beneficial near highways to prevent spillover effects, whereas low stability in specific areas can provide the flexibility needed to alleviate traffic congestion by enabling alternative routes. 
Recognizing path stability's dual role as an indicator of resilience and flexibility enables city planners and transportation authorities to more effectively manage urban mobility, addressing the diverse needs of an evolving urban landscape.

\bigskip

\noindent \textbf{Acknowledgments.} GC has been partially supported by PNRR (Piano Nazionale di Ripresa e Resilienza) in the context of the research program 20224CZ5X4 PE6 PRIN 2022 ”URBAI – Urban Artificial Intelligence” (CUP B53D23012770006), funded by the European Commission under the Next Generation EU programme.
MN has been partially supported by PNRR - M4C2 - Investimento 1.3, Partenariato Esteso PE00000013 - ”FAIR – Future Artificial Intelligence Research” – Spoke 1 ”Human-centered AI”, funded by the European Commission under the NextGeneration EU programme and by the Horizon Europe research and innovation programme under the funding scheme Green.Dat.AI, G.A. 101070416.

\appendix{}

\begin{table}[!h]
\resizebox{\textwidth}{!}{
\begin{tabular}{l l|c c c c c c c c c}
\toprule
\textbf{rank} & \textbf{City} & \textbf{$\mathbf{S_\Delta(o,d)}$} & \textbf{IQR} & \textbf{length} $\mathbf{p(d, d^x)}$ \textbf{[m]} & \textbf{R} & 
\textbf{avg. street length [m]} & \textbf{std street length [m]} & \textbf{avg. circuity} & \textbf{entropy of bearing} & \textbf{Cluster} \\
\midrule
\small{1st} & Vancouver & 0.964 & 0.123 & 104.459 & 0.014 & 143.491 & 187.235 & 1.058 & 2.804 & 2 \\
\rowcolor{lavender!80} \small{2nd} & Buenos Aires & 0.964 & 0.155 & 202.879 & 0.019 & 106.532 & 78.974 & 1.011 & 3.421 & 3 \\
\small{3rd} & Tokyo & 0.967 & 0.094 & 143.515 & 0.010 & 67.088 & 71.896 & 1.049 & 3.526 & 1 \\
\rowcolor{lavender!80} \small{4th} & Karachi & 0.968 & 0.096 & 110.417 & 0.015 & 77.797 & 102.936 & 1.029 & 3.534 & 1 \\
\small{5th} & Athens & 0.969 & 0.115 & 156.017 & 0.017 & 91.089 & 149.196 & 1.052 & 3.535 & 3 \\
\rowcolor{lavender!80} \small{6th} & Lima & 0.970 & 0.100 & 170.522 & 0.017 & 83.527 & 99.988 & 1.046 & 3.522 & 3 \\
\small{7th} & Tehran & 0.972 & 0.111 & 245.429 & 0.021 & 91.523 & 167.629 & 1.057 & 3.408 & 3 \\
\rowcolor{lavender!80} \small{8th} & Cape Town & 0.974 & 0.107 & 115.621 & 0.014 & 106.741 & 151.034 & 1.096 & 3.537 & 1 \\
\small{9th} & Barcelona & 0.974 & 0.129 & 242.051 & 0.024 & 108.123 & 194.674 & 1.091 & 3.493 & 3 \\
\rowcolor{lavender!80} \small{10th} & Dubai & 0.975 & 0.110 & 167.475 & 0.019 & 117.045 & 185.789 & 1.061 & 3.433 & 3 \\
\small{11th} & Osaka & 0.977 & 0.084 & 131.615 & 0.009 & 74.131 & 102.240 & 1.066 & 3.327 & 1 \\
\rowcolor{lavender!80} \small{12th} & Sydney & 0.977 & 0.078 & 134.706 & 0.013 & 124.176 & 157.358 & 1.065 & 3.505 & 1 \\
\small{13th} & Manila & 0.977 & 0.087 & 129.476 & 0.013 & 88.026 & 109.152 & 1.067 & 3.555 & 1 \\
\rowcolor{lavender!80} \small{14th} & Cairo & 0.978 & 0.076 & 141.117 & 0.010 & 74.485 & 122.122 & 1.036 & 3.454 & 1 \\
\small{15th} & Toronto & 0.979 & 0.115 & 101.269 & 0.012 & 161.453 & 164.075 & 1.122 & 2.614 & 2 \\
\rowcolor{lavender!80} \small{16th} & Mumbai & 0.979 & 0.072 & 107.801 & 0.014 & 122.786 & 170.461 & 1.069 & 3.513 & 1 \\
\small{17th} & Melbourne & 0.980 & 0.082 & 140.807 & 0.011 & 93.756 & 130.105 & 1.051 & 2.910 & 2 \\
\rowcolor{lavender!80} \small{18th} & Jakarta & 0.981 & 0.071 & 148.323 & 0.012 & 68.916 & 89.483 & 1.069 & 3.385 & 1 \\
\small{19th} & Mexico City & 0.981 & 0.083 & 148.028 & 0.010 & 86.447 & 119.765 & 1.044 & 3.276 & 1 \\
\rowcolor{lavender!80} \small{20th} & New York City & 0.981 & 0.088 & 163.346 & 0.010 & 130.767 & 126.963 & 1.040 & 3.387 & 1 \\
\small{21st} & Kinshasa & 0.982 & 0.073 & 92.759 & 0.010 & 108.892 & 176.730 & 1.047 & 3.554 & 1 \\
\rowcolor{lavender!80} \small{22nd} & Los Angeles & 0.982 & 0.085 & 112.080 & 0.008 & 151.405 & 146.615 & 1.045 & 3.290 & 1 \\
\small{23rd} & Lagos & 0.984 & 0.062 & 128.778 & 0.010 & 116.709 & 106.864 & 1.050 & 3.563 & 1 \\
\rowcolor{lavender!80} \small{24th} & Beijing & 0.984 & 0.091 & 151.000 & 0.011 & 205.068 & 263.547 & 1.055 & 2.976 & 2 \\
\small{25th} & Istanbul & 0.985 & 0.066 & 120.009 & 0.008 & 88.926 & 140.791 & 1.061 & 3.580 & 1 \\
\rowcolor{lavender!80} \small{26th} & New Delhi & 0.986 & 0.049 & 121.806 & 0.007 & 79.577 & 110.559 & 1.043 & 3.571 & 1 \\
\small{27th} & Seoul & 0.987 & 0.066 & 111.498 & 0.007 & 118.322 & 177.761 & 1.059 & 3.576 & 1 \\
\rowcolor{lavender!80} \small{28th} & Sao Paulo & 0.987 & 0.052 & 146.110 & 0.008 & 105.871 & 132.631 & 1.065 & 3.578 & 1 \\
\small{29th} & Paris & 0.987 & 0.060 & 155.315 & 0.010 & 107.996 & 150.773 & 1.058 & 3.580 & 1 \\
\rowcolor{lavender!80} \small{30th} & Shanghai & 0.988 & 0.089 & 121.100 & 0.011 & 257.970 & 297.162 & 1.041 & 3.485 & 0 \\
\small{31st} & Dallas & 0.988 & 0.052 & 108.245 & 0.007 & 152.761 & 174.535 & 1.061 & 3.305 & 1 \\
\rowcolor{lavender!80} \small{32nd} & London & 0.989 & 0.043 & 106.723 & 0.007 & 104.174 & 148.343 & 1.065 & 3.547 & 1 \\
\small{33rd} & Bangkok & 0.989 & 0.035 & 143.232 & 0.008 & 92.980 & 120.363 & 1.058 & 3.551 & 1 \\
\rowcolor{lavender!80} \small{34th} & Shenzhen & 0.990 & 0.080 & 97.287 & 0.008 & 172.760 & 281.984 & 1.077 & 3.450 & 0 \\
\small{35th} & Guangzhou & 0.990 & 0.064 & 110.308 & 0.008 & 207.269 & 287.831 & 1.066 & 3.488 & 0 \\
\rowcolor{lavender!80} \small{36th} & Boston & 0.991 & 0.049 & 102.397 & 0.006 & 141.867 & 150.132 & 1.080 & 3.566 & 1 \\
\small{37th} & Amsterdam & 0.992 & 0.051 & 93.703 & 0.007 & 110.768 & 232.475 & 1.085 & 3.551 & 0 \\
\rowcolor{lavender!80} \small{38th} & Berlin & 0.992 & 0.037 & 81.845 & 0.006 & 165.183 & 219.260 & 1.050 & 3.556 & 0 \\
\small{39th} & Ottawa & 0.993 & 0.043 & 80.980 & 0.006 & 213.077 & 319.846 & 1.120 & 3.354 & 0 \\
\rowcolor{lavender!80} \small{40th} & Milan & 0.994 & 0.031 & 93.705 & 0.005 & 102.955 & 184.624 & 1.062 & 3.554 & 1 \\
\small{41st} & Rome & 0.994 & 0.035 & 92.208 & 0.006 & 140.145 & 266.040 & 1.083 & 3.578 & 0 \\
\rowcolor{lavender!80} \small{42nd} & Moscow & 0.994 & 0.038 & 86.470 & 0.006 & 217.438 & 268.513 & 1.072 & 3.570 & 0 \\
\small{43rd} & Dhaka & 0.995 & 0.021 & 77.116 & 0.004 & 149.710 & 202.884 & 1.080 & 3.272 & 0 \\
\rowcolor{lavender!80} \small{44th} & Brussels & 0.995 & 0.027 & 81.341 & 0.004 & 159.164 & 207.526 & 1.059 & 3.579 & 0 \\
\small{45th} & Bogota & 0.996 & 0.033 & 81.536 & 0.004 & 110.761 & 226.609 & 1.105 & 3.484 & 0 \\
\rowcolor{lavender!80} \small{46th} & Hamburg & 0.996 & 0.018 & 74.775 & 0.004 & 167.510 & 261.195 & 1.073 & 3.565 & 0 \\
\bottomrule
\end{tabular}}
\caption{Stability and network metrics for 46 cities, ranked by median stability from least to most stable. Metrics include median path stability \( S_\Delta(o,d) \), interquartile range (IQR) of stability, average length of \(p(d, d^x) \), ratio \( R \), average and standard deviation of street length, average circuity, entropy of road bearing, and assigned cluster.}
\label{tab:divercity_single_cities}
\end{table}

\bibliography{sn-bibliography}
\end{document}